\documentclass[aps,superscriptaddress,eqsecnum,nofootinbib,showpacs,preprintnumbers]{revtex4-2}
\usepackage{amsmath}
\usepackage{amssymb}
\usepackage{amsfonts,amsthm,bm}
\usepackage{color,xcolor}
\usepackage[greek,english]{babel}
\usepackage{titlesec}
\usepackage{comment}
\usepackage{appendix}
\usepackage{graphicx,epsfig}
\graphicspath{{figures/}{./}}
\usepackage{float}
\usepackage{subfigure}
\usepackage{tikz}
\newcommand{\abs}[1]{\left| #1 \right|}

\newcommand{\be}{\begin{eqnarray}}
	\newcommand{\ee}{\end{eqnarray}}
\newcommand{\bea}{\begin{eqnarray}}
	
	\newcommand{\eea}{\end{eqnarray}}

\newcommand{\beq}{\begin{equation}}
	\newcommand{\eeq}{\end{equation}}
\newcommand{\bseq}{\begin{subequations}}
	\newcommand{\eseq}{\end{subequations}}
\begin{document}

\preprint{\leftline{KCL-PH-TH/2024-{\bf 07}}}
 
	\title{Condensate-Induced Inflation from Primordial Gravitational Waves in String-Inspired Chern-Simons Gravity}
	
	\author{Panagiotis  Dorlis }
	\email{psdorlis0@gmail.com} 
	\affiliation{Physics Division, School of Applied Mathematical and Physical Sciences,
		National Technical University of Athens,  Zografou Campus,
		Athens 15780, Greece.}
	\author{Nick E. Mavromatos}
	\affiliation{Physics Division, School of Applied Mathematical and Physical Sciences,
		National Technical University of Athens,  Zografou Campus,
		Athens 15780, Greece.}
	\affiliation{Physics Department, King's College London, Strand, London WC2R 2LS, UK.}

	\author{Sotiris-Neilos Vlachos }
	\email{sovlacho@gmail.com}
	\affiliation{Physics Division, School of Applied Mathematical and Physical Sciences,
		National Technical University of Athens,  Zografou Campus,
		Athens 15780, Greece.}

	\vspace{17.5cm}
	\begin{abstract}
	In this work, we elaborate further on a cosmological model of inflation that characterises Chern-Simons (CS) gravity models inspired from string theory. Such models are known to belong to the class of the so-called String-Inspired Running Vacuum Cosmologies. In particular, by applying methods of dynamical systems, commonly used in scalar-field cosmology, we examine in detail, for the first time, the passage from a pre-inflationary era dominated by a stiff-axion-matter equation of state, characteristic of the model, to inflation of Running Vacuum Model (RVM) type. By a careful discussion of the formation of the condensate of the CS gravitational anomaly term, induced by populations of primordial gravitational waves at the end of the stiff-axion-matter era, we show that an effectively linear axion-monodromy potential arises. This eventually causes the transition from the matter to the RVM inflation. By taking into account terms that have previously been ignored in the relevant literature of weak-graviton quantisation, we show that the effect of such terms is to diminish the value of the condensate by half, remaining however in the same order of magnitude. This, in turn, implies that the qualitative conclusions of previous works on the subject remain valid. Moreover, on assuming the approximate cosmic-time independence of the gravitational-CS condensate, we also provide an estimate of the number of sources of the primordial gravitational waves, upon the requirement of respecting the transplanckian conjecture.  
	\end{abstract}
	\vspace{3.5cm}

	\maketitle
	
	\flushbottom
	
	\tableofcontents
 \section{Introduction}

The existence of an early inflationary phase of the Universe, in which the latter undergoes an exponential expansion of its scale factor with the cosmic time,  can explain several of the features of the 
observable Universe. Yet, there is no singly preferred microscopic model of inflation, but rather several models that can fit the inflationary framework~\cite{infl}. The latest cosmological data, especially from Cosmic Microwave Background (CMB) Observations~\cite{Planck:2018vyg}, can rule out several single-inflaton-field models, but several remain. 

From the latter, one of the most optimal models, from the point of view of fitting the data, is the Starobinsky model~\cite{staro}, which is a higher-curvature model, with the higher-order terms arising from a conformal anomaly, and hence their coefficient 
can only be phenomenologically determined. The model is not characterised by a fundamental inflaton field, but a dynamical scalar degree of freedom, responsible for inflation, is hidden in the non-linear gravitational corrections.
Interestingly enough, the Starobinsky model can evade~\cite{Lust} the stringent swampland criteria~\cite{swamp1,swamp2,swamp3,swamp4,swamp5}.\footnote{See, however 
the cautionary remarks of \cite{Kehagias}, supporting the alleviation of the tensions between inflation and the swampland criteria 
in generic slow-roll single-field inflationary models, where the initial conditions for the 
adiabatic curvature perturbations are not known.  Moreover, multifield inflationary models can avoid the swampland criteria.\cite{Achucarro}.} 

Another interesting and compatible with the data cosmological framework, where the inflation is also not due to external inflaton fields but rather to non-linearities of gravity in the early epochs of the Universe, is the so-called running-vacuum model (RVM) of cosmology~\cite{rvm1,rvm2,rvm3,Lima:2013dmf,Lima:2015mca,yu,rvmqft1,rvmqft2,rvmqft3,rvmqft4}, which is phenomenologically consistent and can even contribute to the alleviation of the currently observed cosmological tensions~\cite{rvmdata,tsiapi2,tsiapi,rvmtens}. 

A string-inspired version of the RVM, called Stringy RVM (StRVM) has been proposed in \cite{bms,bms2,bms3,ms1,ms2,Mavromatos:2022xdo,gms}, which is also not in tension with the swampland criteria~\cite{srvmswamp}. This version is a Chern-Simons gravity~\cite{Duncan:1992vz,Jackiw,Alexander:2009tp}, coupled to axionic (pseudo - scalar) degrees of freedom, and the inflationary RVM-like phase arises through the emergence of condensates of the anomalous gravitational Chern-Simons (CS) terms, due to condensation of primordial gravitational waves (GWs)~\cite{bms,ms1,ms2,Mavromatos:2022xdo}. The condensation is a consequence of the fact that, in the presence of chiral GW perturbations, the gravitaional CS terms which couple to the axionic degrees of freedom are non-trivial~\cite{Lue:1998mq,Alexander:2004us,Lyth:2005jf}.

The CS condensate leads to RVM-like inflation, without the necessity for the existence of fundamental inflaton fields. The inflationary epoch is due to gravitational non-linearities that characterise the situation.  It has been shown explicitly~\cite{ms1} that the ground state of the system is characterised by a de-Sitter equation of state, $w=-1$, which is characteristic of an RVM fluid. 
We also note, for completeness, that
an important and necessary aspect of this approach is the existence of a pre-RVM inflationary phase, during which a sufficient amount of primordial GWs are formed, as a result, e.g., of either coalescence of rotating primordial black holes, or non-symmetric collapse of domain walls, that characterise the early epochs of the stringy RVM. The late stages of the pre-RVM inflationary phase is characterised by a stiff-matter equation of state, $w=1$. Such a phase is the result of a dominance of the 
fluid of the massless stringy axionic degrees of freedom that characterise the 
StRVM. On the other hand, in this StRVM framework, it is assumed 
that
near the Big-Bang there exists another (hill-top) inflationary phase~\cite{Ellis:2013zsa}, which arises from condensates of gravitino fields, appearing in the dynamically broken supergravity effective field theory, in the early era of this string universe~\cite{Alexandre:2013iva,Alexandre:2014lla,ms1,ms2}. This first inflation 
is deemed responsible for the isotropy and homogeneity of the cosmic fluid during the axion-stiff and RVM CS-anomaly-condensate-induced inflationary phase of the universe. For details we refer the interested reader to the relevant literature \cite{ms1,ms2,Mavromatos:2022xdo}. 

In the above works, however, we did not discuss the details of the passage from the stiff- to the condensate-RVM-inflationary phase, which is one of the main objectives of the present article. The other main objective is to calculate the gravitational CS condensate itself, going beyond the approximations used in 
\cite{Alexander:2004us,Lyth:2005jf}, which were adopted in \cite{bms,ms1,ms2,Mavromatos:2022xdo}. As we shall demonstrate quantitatively below, going beyond the approximations made in 
\cite{Alexander:2004us,Lyth:2005jf} implies significant corrections to the value of the CS condensate, which actually is diminished by half. Nonetheless, the order of magnitude remains the same, and therefore the qualitative conclusions of the approach of \cite{bms,ms1,ms2,Mavromatos:2022xdo} remain intact. However, the physics details that allow the passage from the stiff-axion-era to the RVM inflationary era contain important novel characteristics, both from a technical and a conceptual point of view, and therefore are worthy of pointing out. 
We also wish to point out that, as discussed already in \cite{ms1,ms2}, our stiff-axion phase is different in origin from other scenarios in the contemporary literature that involve stiff-matter in the early universe~\cite{Zeldovich:1961sbr,Chavanis:2014lra}. In such scenarios the stiff eras occur after inflation, while in our case the stiff-axion era preceeds the RVM inflation. 

The string-inspired gravitational action, which describes the early-epoch StRVM dynamics, after string compactification to (3+1) spacetime dimensions, is given by~\cite{Duncan:1992vz,str1,str2,pol1,pol2}:\footnote{Our conventions and definitions used throughout this work are: signature of metric $(-, +,+,+ )$, Riemann Curvature tensor 
$R^\lambda_{\,\,\,\,\mu \nu \sigma} = \partial_\nu \, \Gamma^\lambda_{\,\,\mu\sigma} + \Gamma^\rho_{\,\, \mu\sigma} \, \Gamma^\lambda_{\,\, \rho\nu} - (\nu \leftrightarrow \sigma)$, Ricci tensor $R_{\mu\nu} = R^\lambda_{\,\,\,\,\mu \lambda \nu}$, and Ricci scalar $R_{\mu\nu}g^{\mu\nu}$. We also work in units $\hbar=c=1$.} 
\begin{align}\label{sea3}
S^{\rm eff (I)}_{\rm B} =&\; \int d^{4}x\sqrt{-g}\Big[ \dfrac{1}{2\kappa^{2}}\, R - \frac{1}{2}\, \partial_\mu b \, \partial^\mu b - \sqrt{\frac{2}{3}} \, \frac{\alpha^\prime}{96\, \kappa} \, b(x) \, R_{\mu\nu\rho\sigma}\, \widetilde R^{\nu\mu\rho\sigma}  + \dots \Big],
\end{align}
where $\kappa^2= M_{\rm Pl}^{-1}$ is the (3+1)-dimensional gravitational coupling, with $M_{\rm Pl} \simeq 2.4 \cdot 10^{18}~{\rm GeV}$ the reduced Planck mass,  $\alpha^\prime =M_s^{-2}$ is the Regge slope, with $M_s$ the string mass scale, and the dots $\dots$ denote gauge as well as higher-derivative terms appearing in the string effective action, that we ignore for our discussion here.  In \eqref{sea3}, the symbol $\widetilde{(...)}$ denotes the corresponding dual tensor, 
defined in section \ref{sec:gwcs} below ({\it cf.} \eqref{dualriem}), and $b(x)$ is the so-called string-model independent axion~\cite{svrcek}, associated with the pseudo - scalar degree of freedom of the massless gravitational ground state multiplet of the superstring (the other members of which are the graviton and dilaton, which has been ignored for our purposes, being assumed constant).

The inflationary dynamics is associated with the formation of a condensate of the gravitational CS anomalous term.\footnote{In this work we follow the conventions for the definition of the gravitational CS term of \cite{kaloper}, which differs by a sign from that of \cite{bms,ms1,ms2,Mavromatos:2022xdo}.} 
\begin{align}\label{cscond}
 \langle R_{CS}\rangle \equiv \frac{1}{2} \, \langle  R_{\mu\nu\rho\sigma}\, \widetilde R^{\nu\mu\rho\sigma} \rangle \,,
\end{align}
which during the inflationary epoch is approximately constant, being a functional of the Hubble parameter $H$, which is approximately constant during inflation. The non-zero result of the condensate is guaranteed in the presence of parity-violating metric perturbations, such as chiral GWs. 
As a function of $H$ the condensate is proportional to $H^4$, which was to be expected on dimensional grounds. As already mentioned, in this paper we shall reevaluate this condensate, relaxing some of the approximations made in \cite{Alexander:2004us,Lyth:2005jf}. In the phase of a non-zero CS condensate, 
the effective action \eqref{sea3}, reduces to that of an axion monodromy, with a linear axion $b$ potential, 
\begin{align}\label{linpotb}
V(b) = {\rm constant} \times b(x) 
\end{align}    
which formally is similar to the situation encountered in 
compactified brane models~\cite{silver}, which leads to linear inflation. The difference of such linear axion models from our condensate case, lies on the fact that in our case the axion potential assumes the form 
\begin{align}\label{H4infl}
V(b) = \mathfrak{d} \, H^4 \, b(x)\,, \quad \mathfrak{d}={\rm constant}, 
\end{align}
where the Hubble parameter is in principle dependent on the cosmic time $t$. Thus, in our condensate case, the resulting cosmic vacuum energy density acquires a conventional RVM form~\cite{rvm1,Lima:2013dmf}, and the induced inflation is precisely due to the non-linearities associated with the cosmic evolution of $H(t)$, and not to a fundamental inflaton field, like the axion $b$ in the linear-axion monodromy models of \cite{silver}. Nonetheless, in the phase of a (approximately) constant condensate, due to the form of the effective action \eqref{sea3}, one can still use the dynamical systems approach to inflation due to scalar fields, in order to study the passage from the stiff-axion into the RVM inflationary epoch in the case of the StRVM, which will be one of the main topics of the current article.\footnote{Our approach is distinct from other approaches based on string-inspired higher-curvature  (eg Gauss-Bonnet) effective actions, in the presence of CS terms coupled to scalar fields, such as those in \cite{Satoh:2007gn,Satoh:2008ck}. Those works also predict detectable in principle frequencies for circular polarisation due to the CS terms.}

The structure of the paper is as follows: in the next section \ref{dynamical_system_approach}, we  describe briefly the dynamical system approach to (pseudo)scalar field cosmology, in which we first express the cosmological equations as an autonomous dynamical system of ordinary differential equations. Then, we consider the case of a linear (pseudo)scalar potential, and show that inflation corresponds to a saddle  point of the evolution, while also that an approximately exponential expansion, with the phenomenologically desired number of e-folds, occurs only for a specific range of initial conditions.
In section \ref{sec:cond}, we explain in detail how the (approximately) linear axion potential arises from condensation of primordial GWs, and evaluate the potential by applying the (weak quantum gravity) method of \cite{Lyth:2005jf}, but going beyond the approximations involved in that work, as well as in \cite{Alexander:2004us}.  We calculate the CS condensate due to primordial GW in the stiff era. Then, in section \ref{sec:infl}, 
we calculate the CS condensate in the inflationary era, and discuss the detailed conditions for a smooth transition from the stiff epoch to the inflationary era. Finally, conclusions and outlook are presented in section \ref{sec:concl}. Some technical aspects of our approach of dynamical systems to cosmology are presented in two Appendices, \ref{appendixA} and \ref{appendixB}.
 
\section{Scalar Field Cosmology - Dynamical Approach}\label{dynamical_system_approach}

In this section, we consider the scalar field cosmology dynamics, by considering  minimal coupling and a scalar field which respects the symmetries of the underlined Friedmann-Lemaitre-Robertson-Walker (FLRW) spacetime geometry. Although the StRVM is associated with pseudo - scalar fields $b(x)$, nonetheless under the assumptions of isotropy and homegeneity, that we are employing for our purposes here,  
the pseudo - scalar nature of the field is not relevant, and thus, for brevity in what follows we concentrate on the scalar case. 

Our goal is to express the Friedman and Klein-Gordon equations, that govern the cosmological evolution, in the form of an autonomous dynamical system of ordinary differential equations (ODE). To illustrate that, we introduce some appropriate (dimensionless) variables, the so called Expansion Normalized (EN) variables as in \cite{Bahamonde_2018}. Then, we are interested in a particular case for the scalar potential. Specifically, we are concerned with a linear potential, for which the analysis of the corresponding phase space shows that inflation arises as a saddle point of the evolution, implying in this way a graceful exit, too. The latter is a necessary condition in order not to spoil the successful predictions of the standard cosmology, such as nucleosynthesis. As we will show in the following, inflation is reached only under a certain class of initial conditions, contrary to the chaotic inflation of the even power potentials \cite{Belinsky:1985zd, Urena-Lopez:2007zal,Reyes-Ibarra:2010jje}, for which inflation corresponds to an attractor of the evolution. Such a class is consistent with a nearly stiff matter dominated pre-inflationary era, which is introduced in a string-inspired cosmological set up \cite{ms1,ms2,Mavromatos:2021hai} and we will try to examine in detail in this article.

\subsection{The cosmological evolution as an autonomous dynamical system}\label{sec:EN}
 
The general action for a minimally coupled interacting scalar field $b(x)$ reads:
\begin{equation}
	S=\int d^4x \,\sqrt{-g}\,  \left[\frac{R}{2\kappa^2}-\frac{1}{2}(\partial_\mu b)(\partial^\mu b) -V(b) \right]\,,
\label{eq:Action_dynamical}
\end{equation}
where $V(b)$ denotes a self-interaction potential for the scalar field, which for our purposes in this work we take to be positive. The gravitational field equations are given by:
\begin{equation}
	G_{\mu\nu}=\kappa^2 T^{b}_{\mu\nu}\,,
\label{eom1}
\end{equation}
where
\begin{equation}\label{stress}
    T^b_{\mu\nu}=\partial_\mu b\partial_\nu b-\frac{1}{2}g_{\mu\nu}(\partial b)^2 - g_{\mu\nu}V(b)
\end{equation}
is the stress energy tensor of the scalar field $b$.

Variation with respect to the field $b(x)$ gives the well-known Klein-Gordon equation in curved spacetime:
\begin{equation}
    \Box b -V_{, b} = 0\,,
\end{equation}
where commas denotes functional differentiation with respect to the field $b(x)$. We consider a spatially flat FLRW spacetime line element, 
\begin{equation}
    ds^2=-dt^2+\alpha^2(t)\delta_{ij}dx^idx^j\,,
\end{equation}
where $\alpha (t)$ denotes the scale factor.

The gravitational equations of motion \eqref{eom1} reduce to the Friedmann equations :
\begin{align}
    3H^2 &= \kappa^2\left(\frac{\dot{b}^2}{2}+V(b)\right)\,,
    \label{freidmann1}\\ 
    2\dot{H} + 3H^2 &= -\kappa^2\left(\frac{\dot{b}^2}{2}-V(b)\right)\,.
    \label{friedmann2}
\end{align}
and the Klein-Gordon equation for the scalar field becomes:
\begin{equation}
    \ddot{b}+3H\dot{b}+V_{,b}=0\,.
    \label{KG_frw}
\end{equation}
The energy density and the pressure for the scalar field fluid are given by:
\begin{align}
    \rho &= \frac{\dot{b}^2}{2}+V(b)\,, \\ 
    p &= \frac{\dot{b}^2}{2}-V(b)\,,
\end{align}
where the equation of state, $p=\omega_b \rho$, is dynamical and reads:
\begin{equation}
\omega_b=\frac{\frac{\dot{b}^2}{2}-V(b)}{\frac{\dot{b}^2}{2}+V(b)}\,,
    \label{eq_of_State}
\end{equation}
which ranges from $+1$ (dominance of kinetic energy), to $-1$ (dominance of potential energy).

At this point, we introduce the EN variables as follows: 
    \begin{equation}
    x=\frac{\kappa \dot{b}}{\sqrt{6}H} \ \ \text{and} \ \  y=\frac{\kappa\sqrt{\abs{V}}}{\sqrt{3}H}\,.
    \label{ENvariables}
\end{equation}

By its definition, the variable $y$ is always non-negative, $y\geq 0$. This provides a by-definition constraint on the physical phase space.  Another constraint of the physical phase space comes from  the Friedmann equation \eqref{freidmann1} (Friedman constraint), which takes the following simple form:
\begin{equation}
    x^2 +y^2=1
    \label{Friednmann_in_EN}\,, 
\end{equation}
thereby constraining the physical phase space on the unit circle of the $x-y$ plane. Thus, the physical phase space on the  $(x, y)$-plane is represented by the positive $y$ half-circle $(y\geq0)$, with the center at the origin. The equation of state \eqref{eq_of_State} becomes:
\begin{equation}
    \omega_b=\frac{x^2-y^2}{x^2+y^2}=x^2-y^2 \,.
\label{eq_of_state_axion}
\end{equation}
What we can see now is that, for $x=0$ and $y=1$, the potential energy of the scalar field dominates, giving an equation of state $\omega=-1$, behaving in this way as an effective cosmological constant which drives the accelerated expansion of the universe. For $x=1$ and $y=0$, the kinetic energy of the field dominates, and we have an equation of state $\omega=1$, where the scalar field behaves as stiff matter. 

With the EN variables, we can finally express the equations \eqref{freidmann1},\eqref{friedmann2} and \eqref{KG_frw} as an autonomous dynamical system of ordinary differential equations (ODE) as follows (see details of these calculations in Appendix \ref{appendixA}):
\begin{align}
    x^{\prime}&=-\frac{3}{2}\left[2x - x^3 +x\left(y^2-1\right)-\frac{\sqrt{2}}{\sqrt{3}}\lambda y^2\right] \label{dyn1}\,,  \\ 
    y^{\prime}&=-\frac{3}{2}y\left[-x^2 + y^2-1 + \frac{\sqrt{2}}{\sqrt{3}}\lambda x\right] \label{dyn2} \,,
    \\ 
    \lambda^{\prime}&=-\sqrt{6}\left(\Gamma -1 \right)\lambda^2 x \,,
    \label{dyn3}
\end{align}
where a prime denotes differentiation with respect to the time parameter $N \equiv log\big(\alpha(t)\big)$ (e-folds), and we have defined:
\begin{equation}
    \lambda=-\frac{V_{,b}}{\kappa V} \ \ \text{and} \ \ \Gamma=\frac{V V_{,bb}}{V^{2}_{,b}}
\end{equation}
One should notice that the above system of ODE is three dimensional (3D) rather than two dimensional (2D), which one would naively expect to be the case of the EN variables. The presence of the extra dimension $\lambda$ is 
necessitated by the requirement of 
autonomy of the system in the general case, since the EN variables are time dependent.

\begin{figure}[ht!]
    \centering
\includegraphics[width=0.5\textwidth]{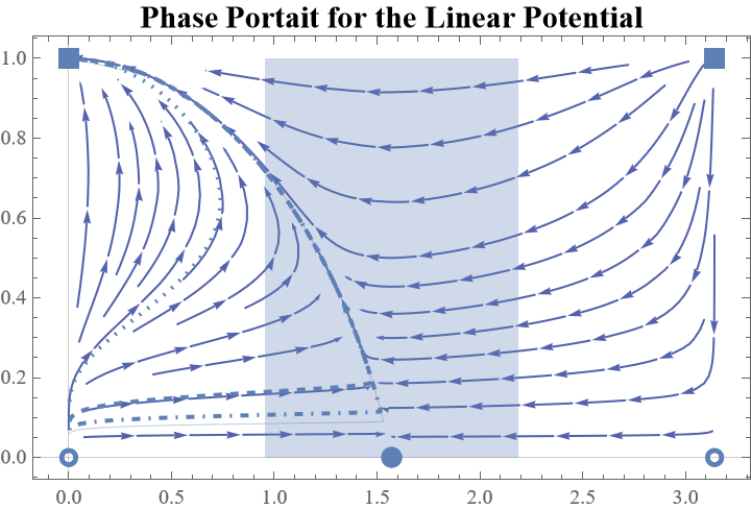}
    \caption{The phase portrait of the dynamical system \eqref{phi_prime},\eqref{z_prime}. The critical points are marked on the Figure as squares, circles or a disc.  The shaded reactangular region corresponds to accelerated expansion, with equation of state $\omega_b\leq-1/3$, with the disk in the center of the abscissa denoting exact exponential expansion, $\omega_b=-1$. The solid, dot-dashed, dashed and dotted  lines correspond to orbits with initial conditions $\varphi_i=10^{-5/2}$ and $\zeta_i=0.06,\;0.07,\;0.09,\;0.14$, respectively.} 
    \label{stream}
\end{figure}

However, the introduction of $\lambda$ can be avoided in specific cases of potentials. One famous example where such a reduction occurs is the  quintessence model with exponential potential \cite{Copeland:1997et,Urena-Lopez:2011gxx}, for which $\lambda$ is just a constant and then acts only as a roll parameter. 
More generally, models in which the only matter field present is the scalar field, such as the StRVM discussed in this work during its 
early phases~\cite{bms,ms1,ms2}, constitute examples where 
the ODE system \eqref{dyn1}-\eqref{dyn3} reduces to a 2D dynamical system. This is a consequence of the Friedman equation \eqref{freidmann1}.\footnote{When matter beyond the scalar field is present, the Friedman equation \eqref{freidmann1} implies $x^2+y^2\leq 1$, i.e. the interior of the unit circle is also part of the phase space and the dynamical system remains 3D. This may happen in realistic StRVM or other string cosmologies, where one may face, e.g., a multiaxion cosmology, due to the presence of axion fields from compacitifcation, in addition to the KR axion. In our context, such axions may be present during the pre-inflationary epoch~\cite{Mavromatos:multiaxion}, thus affecting the details of the passage to the inflationary era.} According to \eqref{freidmann1}, we can express the two variables $(x,y)$ with respect to the angle $\varphi\in [0,\pi]$ of the polar coordinates, as follows:
\begin{equation}
 x=cos\varphi \ ,  \ y=sin\varphi\,.
 \label{cylindrical}
\end{equation}
Then, the two ODE for $(x,y)$ are reduced into one for $\varphi$, and the 2D dynamical system reads:
\begin{equation}\label{ode2}
    \begin{aligned}
      &  \varphi^\prime=\left( 3\cos\varphi -\frac{\sqrt{6}}{2}\lambda   \right) \sin\varphi\,,\\
     &   \lambda^\prime=-\sqrt{6}(\Gamma-1)\lambda^2 \cos\varphi\,.
    \end{aligned}
\end{equation}
In principle, $\lambda$ can take any value on the real axis. However, the symmetry of the system of ODE \eqref{ode2} under the simultaneous transformations $(\varphi\rightarrow \pi-\varphi, \, \lambda\rightarrow -\lambda)$ allows us, without loss of generality, to 
consider only positive values of $\lambda
\ge 0$. In such cases, one can bound the phase space through the following change of variable:
\begin{equation}
    \zeta= \frac{\lambda}{\lambda+1}\,,
\end{equation}
which takes on values in the region $\zeta\in[0,1)$, for $\lambda\in[0,+\infty)$. Thence, the system of ODE reads:
\begin{align}
      &  \varphi^\prime=\left( 3\cos\varphi -\frac{\sqrt{6}}{2}\frac{\zeta}{1-\zeta}   \right) \sin\varphi\,,
      \label{phi_prime}
      \\
      & \zeta^\prime=-\sqrt{6}(\Gamma-1)\zeta^2 \cos\varphi
      \label{z_prime}\,.
    \end{align}
and the phase space is restricted to the interior of a finite rectangle in the $\zeta-\varphi$ plane (see figure \ref{stream}).



\subsection{Linear potential : Inflation as a saddle point of the evolution}\label{linear_potential}

Consider the case of a linear potential for the scalar field,
\begin{equation}
    V(b)= \mathcal{C}b>0\,,
    \label{linearpotential}
\end{equation}
where $\mathcal{C}$ is a constant. The dynamical parameter $\lambda$ and consequently $\zeta$ and $\Gamma$ of the dynamical system \eqref{phi_prime},\eqref{z_prime} read: 
\begin{equation}
    \lambda=-\frac{1}{\kappa b} \ ,   \ \zeta=\frac{1}{1-\kappa b} \ ,  \   \Gamma=0\,,
    \label{zetalinear}
\end{equation}
and consequently, we have the following system of ODE: 
\begin{align}
      &  \varphi^\prime=\left( 3\cos\varphi -\frac{\sqrt{6}}{2}\frac{\zeta}{1-\zeta}   \right) \sin\varphi
      \label{phi_prime_linear}\,,
      \\
      & \zeta^\prime=\sqrt{6}\,\zeta^2 \cos\varphi\,,
      \label{z_prime_linear}
\end{align}
where $\zeta \in  \left[0,1\right)$ and $\phi \in  \left[0,\pi \right]$. 

In Table \ref{full_table_dyn_system},
we can see the critical points of interest $(\phi^\prime=\zeta^{\prime}=0)$, together with the equation of state parameter \eqref{eq_of_state_axion} which is given by $\omega_b=x^2 - y^2={\rm cos}^{2}\phi-{\rm sin}^{2}\phi$ and also the stability of the relative points (See Appendix \ref{appendixB} for details on the stability analysis). The critical points O,I,C lie on the $\zeta=0$ line of the aforementioned finite rectangle on the $\phi-\zeta$ plane, i.e. the phase space region $\phi \in \left[0,\pi\right]$ and $\zeta \in \left[0,1\right)$.

\begin{table}[ht!]
\centering
 \begin{tabular}{||c c c c c c||} 
 \hline
Point & Eigenvalues & Eigenvectors & EoS $\omega_b$ & (Non)Hyperbolic & Stability \\ [0.5ex] 
 \hline\hline
$O(0,0)$ & $\lambda_{1,2}=3,0$ & $\left(
\begin{array}{c}
 1  \\
 0 \\
\end{array}
\right) , \left(
\begin{array}{c}
 0  \\
 1 \\
\end{array}
\right)$ & 1 & Non-hyperbolic & Non-stable  \\

 \hline
 
 $I(\frac{\pi}{2},0)$ & $\lambda_{1,2}=-3,0$ & $\left(
\begin{array}{c}
 1  \\
 0 \\
\end{array}
\right) , \left(
\begin{array}{c}
 \frac{-1}{\sqrt{6}}  \\
 1 \\
\end{array}
\right)$ & -1 & Non-hyperbolic & Saddle  \\

 \hline

 $C(\pi,0)$ & $\lambda_{1,2}=3,0$ & $\left(
\begin{array}{c}
 1  \\
 0 \\
\end{array}
\right) , \left(
\begin{array}{c}
 0  \\
 1 \\
\end{array}
\right)$ & 1 & Non-hyperbolic & Non-stable \\
 \hline
 \end{tabular}
 \caption{In this Table, we present the eigenvalues, eigenvectors and equation of state parameter $\omega_b$ for the field $b$, and  indicate the status of the hyperbolicity and stability properties of the various critical points $O,I,C$ of our dynamical system.}
 \label{full_table_dyn_system}
\end{table}

\begin{figure}[ht!]
    \centering
    \includegraphics[width=0.57\textwidth]{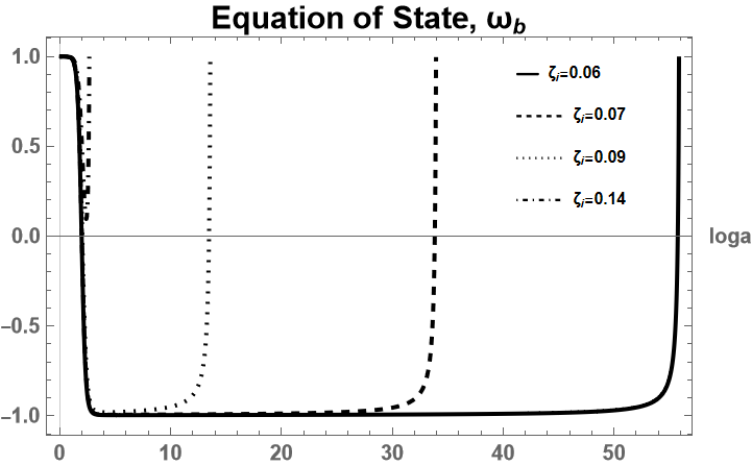}
    \caption{The evolution of the equation of state for the orbits of Figure \ref{stream}. In the case of $\varphi_i=10^{-5/2}$, inflation with the desired number of e-folds, $N>50$, is achieved for $\zeta_i< 0.06$.} 
    \label{eos}
\end{figure}\par 
In Figure~\ref{stream}, the phase portrait of the system is presented. The critical points $O(0,0)$ and $C(\pi,0)$ correspond to an equation of state $\omega_b=+1$ ({\it cf.} \eqref{eq_of_state_axion}), representing a state where the kinetic energy of the axion dominates heavily over the potential energy ("stiff era" phase for the field). The point $I=(\frac{\pi}{2},0)$, which, as shown in Table \ref{full_table_dyn_system} and proved analytically in Appendix~\ref{appendixB}, is a saddle point, corresponds to an equation of state $\omega_b=-1$, i.e. of de-Sitter type. Thus, in this picture, the inflationary state of the universe in our model is represented as saddle critical point of the cosmic evolution. The time that the system stays in the neighborhood of this saddle point is strongly related to the initial conditions of the problem. In Figure \ref{stream}, the $\varphi$ - axis denotes the direction of the eigenvector with eigenvalue $-3$ (stable direction) of the inflationary point $I$. So, only along this direction the point $I=(\frac{\pi}{2},0)$ presents stability. Otherwise, upon small perturbations, the system will only stay in the neighborhood of the saddle point 
for some finite time, and then decay to another phase. The linear potential has the property to provide a saddle point inflationary solution, which is important for a graceful exit from the accelerated phase of the universe. The initial conditions of such a system play a crucial role for the physical outcomes, since the inflationary state cannot be approached for all the possible range of the initial conditions. 
This is visualized by the orbits of 
Figure~\ref{stream}, where we see that there are initial conditions that do not lead to an accelerated expansion at all. From the phase portrait of Figure \ref{stream}, we see that the initial conditions which  lead to an inflationary era are those close to the $\zeta=0$ line of the diagram.  Moreover, in Figure \ref{eos} the evolution of the equation of state for the different orbits reveals that for a given initial value, namely\footnote{The choice of such an initial condition for the $\varphi$ variable will be clarified later in the article.} $\varphi_i=10^{-5/2}$, there is an upper limit $\zeta_i<0.06$ in order for 
the system to acquire an approximately de Sitter phase with the desired number of e-folds, $N>50$. In this sense, the class of the initial conditions that lead to inflation with the desired properties is approximately defined. \par
\begin{figure}
    \centering
\includegraphics[width=0.46\textwidth]{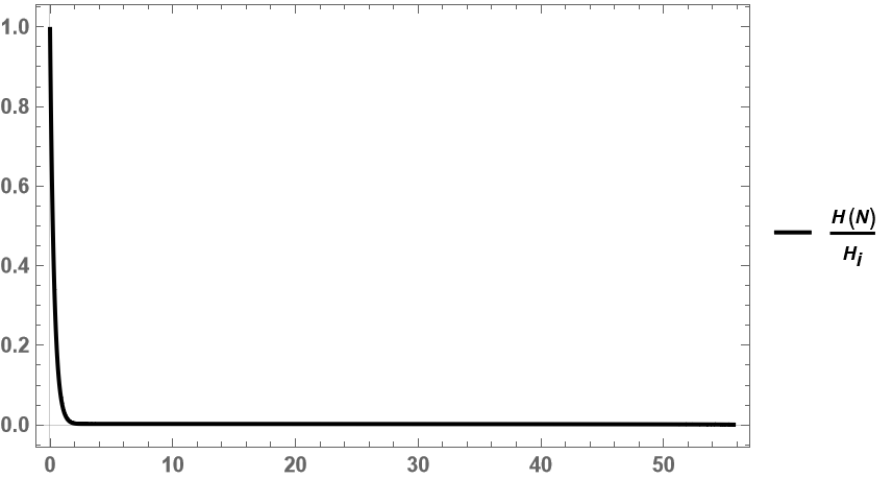}\hfil\includegraphics[width=0.398\textwidth]{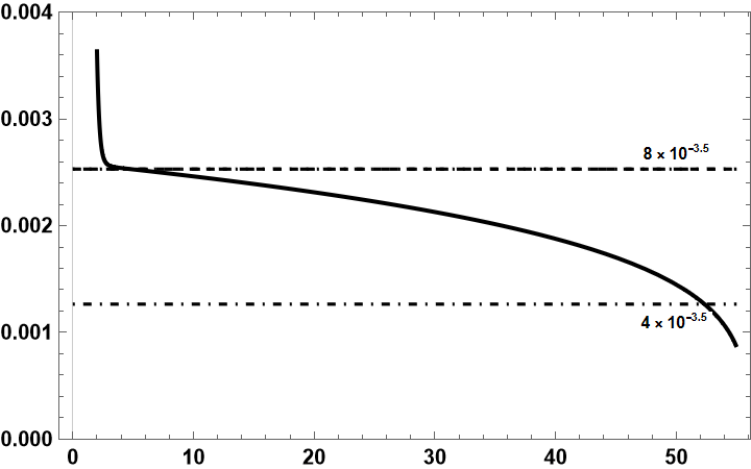}
    \caption{The relative change of the Hubble rate with respect to its initial value at the beginning of the evolution. The Hubble rate during the approximately de Sitter phase has a value which is four orders of magnitude lower than initially, $H_I/H\sim10^{-4}.$}
    \label{HubbleEvolution}
\end{figure}

Supposing that the initial conditions are those for which inflation is achieved, we can have  a relative estimate of the order of magnitude for the Hubble rate with respect to its initial value. Using \eqref{ENvariables} together with \eqref{cylindrical} and \eqref{zetalinear}, we get for the relative evolution of the Hubble rate:
\begin{equation}
 \frac{H(N)}{H_i}= \sqrt{\frac{1/\zeta-1}{1/\zeta_i-1}}\frac{\sin\varphi_i}{\sin\varphi}\,,
\end{equation}
where the subscript $i$ denotes initial values. Choosing the specific initial conditions 
\begin{align}\label{incond|}
(\varphi_i=10^{-5/2},\, \,\zeta_i=0.06)\,, 
\end{align}
we plot the corresponding graph in Figure~\ref{HubbleEvolution}. The Hubble rate decreases very fast, until it acquires an approximately constant value (approximately de Sitter phase), $H_I\approx {\rm constant}$, for $N>50$ e-folds, with an order of magnitude, $H_I\sim 10^{-3.5}H_i$.   

The same procedure can be followed in the numerical evaluation of the evolution of the field $b$ and its time derivative $\dot{b}$. From the definition of the EN variable \eqref{ENvariables},\eqref{cylindrical} and from eq.\eqref{zetalinear}, we obtain:
\begin{equation}
    \frac{b}{M_{\rm Pl}}=\frac{\zeta - 1}{\zeta} \ , \ \ 
     \frac{\dot{b}}{H M_{\rm Pl}}= \sqrt{6} \cos\phi \,.
 \end{equation}

\begin{figure}[h!]
    \centering
\includegraphics[width=0.458\textwidth]{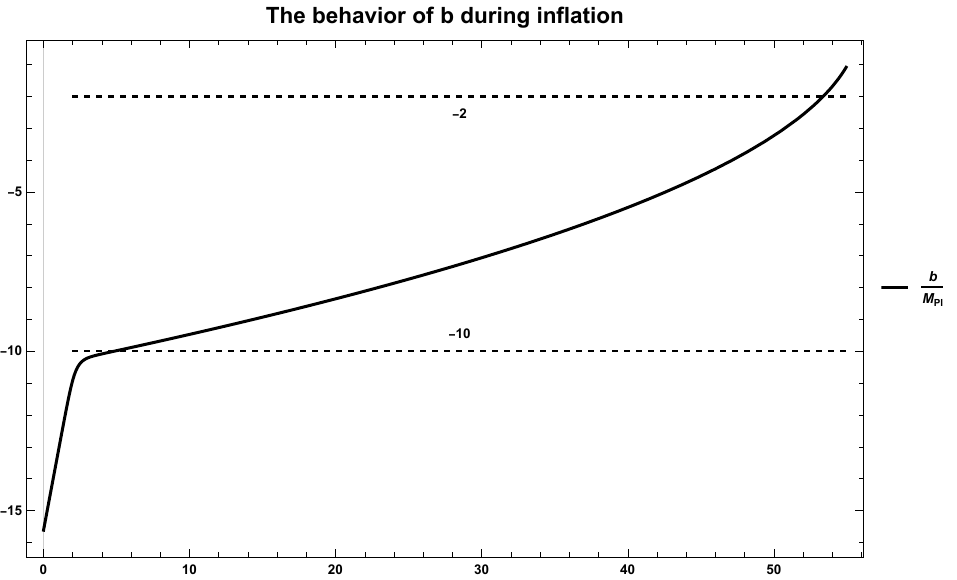}\hfil\includegraphics[width=0.462\textwidth]{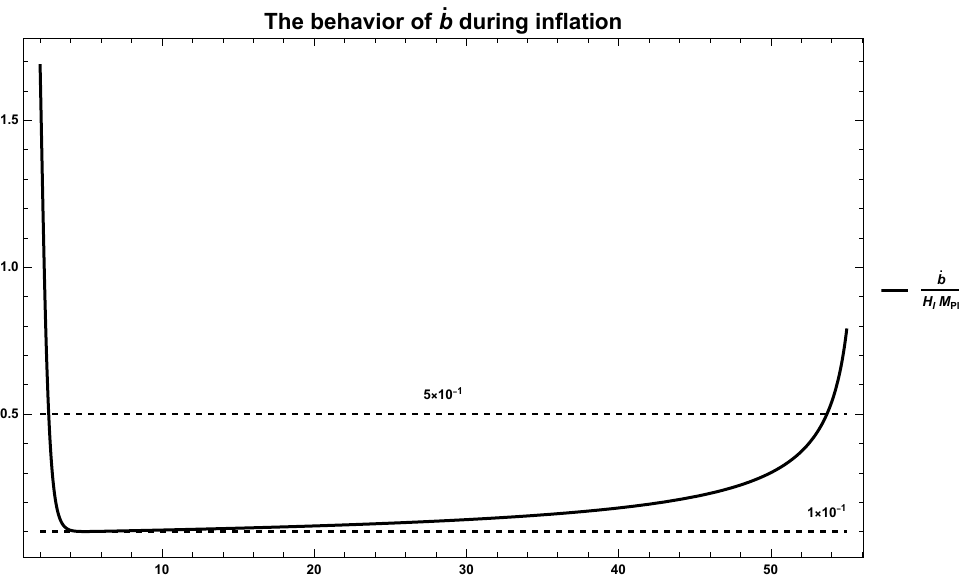}
    \caption{Left panel: numerical evaluation of the behavior of the $b$-field during the dynamical evolution of the Universe, showing that the order of magnitude of $b$ during inflation remains approximately constant. Right panel: the behavior of $\dot{b}$ which also remains in the same order of magnitude during inflation, causing an approximately linear time dependence for the $b$-field.}
    \label{b_and_bdot_infl_fig}
\end{figure}

Considering the same initial conditions \eqref{incond|}, we can plot the result, which is depicted in Figure \ref{b_and_bdot_infl_fig}. We observe that the order of magnitude of both $b$ and $\dot{b}$ does not change during inflation, yielding: 
\begin{equation}
    \frac{\abs{b}}{M_{\rm Pl}} \sim \mathcal{O}\left(10\right) , \ \  \frac{\dot{b}}{H_I M_{\rm Pl}} \sim \mathcal{O}\left(10^{-1}\right).
    \label{b_bdot_orders}
\end{equation}
It is noteworthy to mention here that the fact that the evolution tends to an approximately de Sitter solution, rather than an anti-de Sitter one, is due to the restriction of $V(b)>0$. On the other hand, the fact that the system tends to acquire a cosmological constant type equation of state does not depend on the sign of the potential. The epoch characterised by an approximate equation of state $\omega_b\approx -1$ corresponds to an era of an effective-cosmological-constant dominance, 
\begin{equation}
\Lambda_{eff}=\mathcal{C} b \simeq \rm constant~ .
\label{effective_cosmological_constant}
\end{equation}
since $b$ is approximately constant up to order of magnitude. Moreover, in this epoch $b$ is negative in its sign, which means that whether the evolution is approximately de-Sitter or anti de-Sitter depends on the sign of $\mathcal{C}$. Specifically, inflation corresponds to a negative sign, $\mathcal{C}<0$.  As we shall discuss later on, this is also in agreement with the considerations of \cite{bms,ms1,ms2,Mavromatos:2022xdo}, within the StRVM framework.

\section{Producing The Linear Potential Through Gravitational Condensate}\label{sec:cond}

In what follows, we present a specific way 
by means of which an approximately linear axion potential is generated in the presence of gravitational waves~\cite{Lyth:2005jf,Alexander:2004us} in CS gravity~\cite{Jackiw,Alexander:2009tp}.
This has been proposed in the context of the StRVM in \cite{bms,ms1,ms2,Mavromatos:2022xdo,Mavromatos:multiaxion} 
to discuss RVM inflation. In this section we shall provide a further, and more detailed analysis, on the evaluation of such condensates, starting from the pre-RVM-inflationary stiff-axion-matter era of the model, something that has not been examined before in the relevant literature.\footnote{We stress that in the StRVM context, the stiff era is dominated by massless KR axion matter and occurs before RVM inflation. The possibility of having generic stiff-matter dominance in the early universe has also been discussed in the literature~\cite{Zeldovich:1961sbr,Chavanis:2014lra} but there such epochs occur after inflation, and are due to different mechanisms than the one provided by the StRVM. For instance, in \cite{Zeldovich:1961sbr} the stiff era is due to a cold gas of baryons. In the cases examined in 
\cite{Chavanis:2014lra}, the stiff era can occur in situations where dark matter is made of relativistic self-gravitating Bose-Einstein condensates (BEC), with positive or negative energy densities. It has been shown in that work that in cases of stiff-BEC with negative energy density, the primordial universe may be singularity free. In our StRVM, the pre-inflationary epoch which is characterised by a 
positive energy density stiff-KR-axion matter is preceded~\cite{ms1,ms2} by appropriate first-inflation eras occurring in a phase of dynamically broken supergravities, which appear as the low energy limit of strings. Generically in string theories, the higher curvature corrections that characterise the effective gravitational actions could provide a resilution of the initial singularity~\cite{Antoniadis:1993jc}, independently of the properties of the stiff matter.}
In particular,  
we shall consider a cosmological (FLRW) background during the pre-inflationary stiff era phase of the universe, and show that there is a creation of such a linear potential from the condensation of chiral gravitational waves at the end of the stiff era. This can then lead to an approximately inflationary state which is dynamical and as such of running vacuum type (RVM). As it becomes clear from the previous discussion, whether such an approximate de-Sitter phase is achieved depends on the initial conditions, which in this case corresponds to whether the condensate is created. Thereafter, the dynamics are governed by the effective linear potential for the axion. 

The CS gravitational theory~\cite{Jackiw,Alexander:2009tp} introduces a linear coupling of the (pseudo)scalar field, $b$, with the gravitational CS (gCS) term, $\mathcal{L}_{int}\sim b R_{CS}$, where $R_{CS}$ is given in \eqref{RCS}, and is a total derivative.\footnote{We should stress at this point that the gCS term, unlike the gauge CS term, which the axion also couples to, yields a non trivial variation with the metric tensor, namely the Cotton tensor ({\it cf.} \eqref{cottdef}, below)~\cite{Jackiw}. This implies  the non conservation of the naive axion stress tensor \eqref{stress}, which physically is interpreted as indicating a non trivial exchange of energy between the axion matter and the gravitational field.} The latter is a CP violating term and as such vanishes for spherically symmetric or isotropic and homogeneous spacetime backgrounds. This is the case when an FRW background is considered, which means that in such a case the modified Lagrangian is equivalent to the Einstein - Hilbert of General Relativity. Consequently, the scalar field is minimally coupled to gravity and the cosmological evolution is governed by a stiff equation of state, in which the theory is shift symmetric, i.e. invariant under global transformations of the scalar, 
\begin{align}\label{shiftsym}
b\rightarrow b+ {\rm constant}\,. 
\end{align}
However, when gravitational waves are produced, through non-spherically symmetric coalescence of primordial black holes or collisons of domain walls, the gCS term becomes non-trivial because different helicities of the tensorial perturbations propagate in a different way~\cite{Lue:1998mq,Alexander:2004wk,Alexander:2004us,Lyth:2005jf}. This difference to the wave equations for the left and right - handed polarizations comes through the Cotton tensor \eqref{cottdef}, which, as already mentioned, modifies the gravitational equations leading to gravitational wave birefringence of cosmological origin. 

In this article, we treat the (weak) gravitational waves quantum mechanically by applying the process of second quantization~\cite{Lyth:2005jf}, i.e. by promoting the perturbations to operators through the definition of the corresponding creation and annihilation operators. In this sense, the gCS term becomes also an operator, $\widehat{R}_{CS}$, which we calculate up to second order in the tensorial perturbations. The gCS operator backreacts onto the effective Lagrangian through its vacuum expectation value (vev), $\langle  R_{CS}\rangle \equiv \langle 0|\widehat{R}_{CS}|0\rangle$, where $|0\rangle$ denotes the appropriate gravitational ground state of the system, and the symbol $\widehat{(...)}$ is used to denote quantum operators. We claim that 
$\langle R_{CS}\rangle$
acquires a constant value, thus having the form of a (translationally invariant) gravitational condensate. In this sense, a linear potential for the axion arises, $V_{eff}\sim \langle R_{CS}\rangle b$, ending the stiff-axion-matter dominance, while simultaneously  breaking the shift symmetry \eqref{shiftsym} of the effective gravitational theory. Thus, the formation of the condensate, at the late phase of the stiff era, coincides with the beginning of a cosmic evolution governed by a linear axion potential, studied in section \ref{linear_potential}. From our previous discussion, then, 
it follows that the formation of the condensate has to occur at a phase of the stiff era for which the initial conditions for the linear-potential-governed evolution belong to the desired class for inflation to occur. This is in accordance with the breaking of the shift symmetry of the theory, since the initial conditions, $(\varphi_i,\zeta_i)$, and consequently the subsequent cosmic evolution,  depend on the value of the axion itself and not only on its derivatives.

\subsection{Gravitational Waves in Chern-Simons Gravity}\label{sec:gwcs}

We consider the Chern-Simons Gravitational theory, given by the action \eqref{sea3}:
\begin{equation}
	S=\int d^4x \,\sqrt{-g}\,  \left[\frac{R}{2\kappa^2}-\frac{1}{2}(\partial_\mu b)(\partial^\mu b) - A\, b\,R_{CS}\right]\,,
\label{eq:Action}
\end{equation}
where,
\begin{equation}\label{Aval}
    A=\sqrt{\frac{2}{3}}\frac{\alpha^\prime}{48\kappa}
\end{equation}
denotes the coupling constant of the string-inspired model~\cite{Duncan:1992vz}, which the StRVM is based upon~\cite{bms,ms1,ms2}.

The quantity $R_{CS}$ is the gCS anomaly term:
\begin{equation}
	\label{RCS}
	R_{CS}= \frac{1}{2}R^{\mu}_{\,\,\,\nu\rho\sigma}\widetilde{R}^{\nu\,\,\,\,\rho\sigma}_{\,\,\,\mu},
	\end{equation}
with the symbol $\widetilde{(\dots)}$ denoting the dual of the Riemann tensor, defined as
\begin{equation}
\label{dualriem}
\widetilde{R}_{\alpha\beta\gamma\delta}=\frac{1}{2}R_{\alpha\beta}^{\,\,\,\,\,\,\,\,\rho\sigma}\varepsilon_{\rho\sigma\gamma\delta}\,,
\end{equation}
and $\varepsilon_{\rho\sigma\kappa\lambda} = \sqrt{-g(x)} \, \hat \epsilon_{\rho\sigma\kappa\lambda} $  the covariant Levi-Civita  under the convention that the symbol $\hat \epsilon_{0123}=1$, {\it etc}.\par 
Variation of the action with respect to the metric  and the axion field yields the following equations of motion~\cite{Duncan:1992vz,Jackiw}
\begin{align}
	\label{grav}
	&G_{\mu\nu}=\kappa^2 T^{(b)}_{\mu\nu}+4 \kappa^2 A C_{\mu\nu}~,\\
	\label{Axion}
	&\square\,b=A \, R_{CS}~,
\end{align}
where $T^{(b)}_{\mu\nu}$ is the  stress energy-momentum tensor associated with the kinetic term of a matter field,
\begin{equation}\label{stressb}
	T^{(b)}_{\mu\nu}=\nabla_\mu b\nabla_\nu b-\frac{1}{2}g_{\mu\nu}(\nabla b)^2~.
\end{equation}
The quantity
$C_{\mu\nu}$ is the Cotton tensor derived from the metric variation of $b R_{CS}$ and reads 
\begin{equation}\label{cottdef}
	C_{\mu\nu}=-\frac{1}{2}\nabla^{\alpha}\left[(\nabla^{\beta} b) \widetilde{R}_{\alpha\mu\beta\nu}+(\nabla^{\beta} b) \widetilde{R}_{\alpha\nu\beta\mu}\right]~.
\end{equation}
We assume a spatially flat FLRW spacetime of the following form:
\begin{equation}
    ds^2=-dt^2+\alpha^2(t)\delta_{ij}dx^idx^j \,,
\end{equation}
where $\alpha(t)$ denotes the scale factor. It is easy to obtain that $R_{CS}$ identically vanishes for an FLRW spacetime and the above action describes a massless, real scalar field minimally coupled with gravity, with a stiff equation of state. \par 
However, this is not true in the presence of gravitational waves. The tensor perturbation of the FLRW has the following form: 
\begin{equation}
    ds^2= -dt^2 + \alpha^2(t) (\delta_{ij}+h_{ij})dx^idx^j \,.
    \label{PerturbedFRW}
\end{equation}
One is able to treat the gravitational waves as quantum perturbations, for which the wave equation is derived from the gravitational equations of motion. We can express $h_{ij}$ in the linear polarization basis, expressed as:
\begin{equation}
    h_{ij}=h_+ \epsilon^{(+)}_{ij} + h_\times \epsilon^{(\times)}_{ij} \,,
\end{equation}
where the polarization tensors are defined through:
\begin{align}
  &  \epsilon_{ij}^{(+)}= [e_1(\vec{k})]_i[e_1(\vec{k})]_j-[e_2(\vec{k})]_i[e_2(\vec{k})]_j \,,\\
  &\epsilon_{ij}^{(\times)}= [e_1(\vec{k})]_i[e_2(\vec{k})]_j+[e_1(\vec{k})]_j[e_2(\vec{k})]_i \,,
\end{align}
where $(e_1(\vec{k}),e_2(\vec{k}),e_3(\vec{k}))$, with $e_3(\vec{k})=\vec{k}/\vert\vec{k}\vert$ form a right-handed orthogonal triad of unit vectors. Without loss of generality, we choose the $z-axis$ as the direction of propagation, $e_1=(1,0,0),\;e_2=(0,1,0),\;e_3=(0,0,1)$ and then, $h_{ij}$ reads: 
\begin{equation}
    [h_{ij}]= \begin{bmatrix}
            h_{+} & h_{\times} & 0\\
            h_{\times} & -h_{+} & 0\\
            0 & 0 &0
            \end{bmatrix}\,,
    \label{perturbationmatrix}
\end{equation}
which is traceless $h=h^i_i=0$ and symmetric,  while also $h_{+,\times}=h_{+,\times}(t,z)$. \par

In our analysis below we assume {\it weak} gravitational-wave perturbations. This will prove sufficient for our purposes in this work, which was also the assumption in \cite{Alexander:2004us,Lyth:2005jf,bms,ms1,ms2}. 
In order to obtain the corresponding action of the field theory, one has to expand the Lagrangian of the gravitational theory up to second order in the relevant perturbations, in order to obtain the first order equations of motion through the variational principle. On the other hand, the equations of motion can be obtained directly from the equations of motion of the underlined gravitational theory, by expanding them up to first order, without an explicit reference to the corresponding Lagrangian. Following the latter way, we obtained the equation of motion for each polarization by linearising the gravitational equations of motion \eqref{grav} with respect to $h_{+,\times}$, which leads to the following equations of motion for the gravitational-wave perturbations:
\begin{equation}
    \square h_{\times,+} = \mp \frac{4A\kappa^2}{\alpha^2}\left( 2\dot{\alpha}\dot{b}+\alpha\ddot{b} \right)\partial_t\partial_z h_{+,\times} \mp \frac{4A\kappa^2\dot{b}}{\alpha}\partial_t^2\partial_zh_{+,\times}\pm \frac{4A\kappa^2\dot{b}}{\alpha^3}\partial_z^3h_{+,\times} \,,
    \label{eqh}
\end{equation}
where 
\begin{equation}
    \square = -\partial^{2}_t - 3\frac{\dot{\alpha}}{\alpha}\partial_t+\frac{1}{\alpha^2}\partial^{2}_z\;,
\end{equation}
is the D'Alembertian in the FLRW spacetime. 
These equations imply that the two polarizations of the gravitational field are coupled with each other, due to the non-vanishing contribution of the CP violating coupling of the scalar field with the gCS term. However, the perturbation \eqref{perturbationmatrix} can be expanded according to the helicity basis tensors $\epsilon_{ij}^{L,R}$, as~\cite{Lyth:2005jf}:
\begin{equation}
    h_{ij}(t,\Vec{x})=h_L \ \epsilon_{ij}^{(L)}+h_R \ \epsilon_{ij}^{(R)}=\sum_{\lambda=L,R} h_\lambda(t,\vec{x})\epsilon_{ij}^{(\lambda)}\,,
    \label{helicityexpansion}
\end{equation}
where:
\begin{equation}
  \left [ \epsilon_{ij}^{(R)}\right]=\frac{1}{\sqrt{2}}\left( \left[\epsilon^{(+)}_{ij}\right]+ i \left[\epsilon^{(\times)}_{ij} \right] \right)=
  \frac{1}{\sqrt{2}}\begin{bmatrix}
               1 & i & 0\\
                i & -1 & 0\\
            0 & 0 &0
              \end{bmatrix}=\left[ \epsilon_{ij}^{(L)}\right]^\dagger
              \label{helicitybasis}\,,
\end{equation}
with the polarization tensors obeying the following normalization:
\begin{equation}
\epsilon^{*(\lambda)}_{ij}\epsilon_{(\lambda^\prime)}^{ij}=2\delta_{\lambda\lambda^\prime}\,.
\end{equation}

In the helicity basis, the $R_{CS}$ term has the following structure~\cite{Alexander:2004us,Lyth:2005jf}:
\begin{equation}
    R_{CS}=\frac{1}{2}R^{\mu}_{\,\,\,\nu\rho\sigma}\widetilde{R}^{\nu\,\,\,\,\rho\sigma}_{\,\,\,\mu}=\frac{2 i }{\alpha^3}\left[\left(\partial^{2}_z h_L \partial_z \partial_t h_ R+ \alpha^2 \partial^{2}_t h_L \partial_z \partial_t h_R +\alpha\dot{\alpha} \partial_t h_L \partial_z \partial_t h_R \right) - L \leftrightarrow R \right]+\mathcal{O}(h^4)\,.
\end{equation}
If $h_L$ and $h_R$ happen to satisfy the same dispersion relations, $R_{CS}$ would vanish identically. The only way for $R_{CS}$ to survive is within the existence of a phenomenon called "cosmological birefringence", a prediction about the rotation of the polarization plane of the fields as they travel over cosmological distances. As we stated, this is the case in the presence of tensor perturbation of the FLRW metric, and the $R_{CS}$ survives exactly because $h_R$ and $h_L$ evolve with a difference in the sign of their equations of motion as it is shown in \eqref{eqsofmotions}. 
 Responsible for this are the first-order contributions of the Cotton tensor  in \eqref{grav}, which alter the wave equation of the gravitational waves in a non-trivial way. Due to the form of the Cotton tensor \eqref{cottdef}, the pertinent correction terms contain only derivatives of the axion field, and at least first order derivatives with respect to the perturbations, thereby representing higher-order contributions in the momenta $\vec k$ (with magnitude $k$) of the corresponding Fourier modes.\footnote{When the perturbations are expressed in terms of their Fourier expansions, partial derivatives $\partial_z$ of the perturbations yield powers of the momentum scale $k =|\vec k|$ in the mode expansion, where $\vec k$ 
 is the momentum vector along the direction of propagation of the gravitational waves, which here has been chosen for convenience to be along the $z$ axis, as mentioned above.}
 
 In the helicity basis, therefore, the wave equations \eqref{eqh} take the following form, which decouples the left-handed from the right-handed polarizations: 
\begin{equation}
     \square\, h_{R,L} = \pm \frac{4 i A\kappa^2}{\alpha^2}\left( 2\dot{\alpha}\dot{b}+\alpha\ddot{b} \right)\partial_t\partial_z h_{R,L} \pm \frac{4 i A\kappa^2}{\alpha}\dot{b}\partial_t^2\partial_z h_{R,L} \mp \frac{4 i A\kappa^2}{\alpha^3}\dot{b}\partial_z^3 h_{R,L} \,.
     \label{eqsofmotions}
\end{equation}
Transforming equation \eqref{eqsofmotions} to the conformal time via $dt=\alpha d\eta$, we finally obtain: 
\begin{equation}
h_{\lambda}^{\prime\prime}+2\frac{\alpha^\prime}{\alpha}h_{\lambda}^\prime - \partial^2 _z h_{\lambda}= -l_\lambda \frac{4 i A \kappa^2}{\alpha^2}\partial_z \left(b^{\prime\prime}h_{\lambda}^{\prime}+b^{\prime}h_{\lambda}^{\prime\prime}-b^{\prime}\partial^{2}_z h_{\lambda} \right),\;\;\lambda=R,L\,,
\label{conformaleq}
\end{equation}
where $l_R=+1$ and $l_L=-1$, and a prime denotes differentiation with respect to the conformal time, $\eta$.
Going onto Fourier modes:
\begin{equation}
h_\lambda(\eta,\Vec{x})= \int \frac{d^3k}{(2\pi)^{3/2}} e^{i\vec{k}\cdot{\vec{x}}} \widetilde{h}_{\lambda,\vec{k}}(\eta)\,,
    \label{FourierTans}
\end{equation} 
and substituting onto \eqref{conformaleq}, yields:
\begin{equation}
\widetilde{h}_{\lambda,\Vec{k}}^{\prime\prime}+2\frac{\alpha^\prime}{\alpha}\widetilde{h}_{\lambda,\Vec{k}}^\prime +k^2 \widetilde{h}_{\lambda,\Vec{k}}=  l_{\lambda} l_{\vec{k}}\frac{4 k A \kappa^2 }{\alpha^2} \left(b^{\prime\prime}\widetilde{h}_{\lambda,\Vec{k}}^{\prime}+b^{\prime}\widetilde{h}_{\lambda,\Vec{k}}^{\prime\prime}+ k^2 b^{\prime}\widetilde{h}_{\lambda,\Vec{k}} \right)
\label{general_eq_for_h}\,,
\end{equation}
where  $l_{\vec{k}}=1$ and $l_{-\vec{k}}=-1$. From the definition of the circular polarizations \eqref{helicityexpansion} and \eqref{helicitybasis}, we know that $h_L^*(\eta,\vec{x})=h_R(\eta,\vec{x})$, which for the mode expansion implies
$\widetilde{h}_{L,-\vec{k}}^*(\eta)=\widetilde{h}_{R,\vec{k}}(\eta)$. As such, $\widetilde{h}_{L,-\vec{k}}^*(\eta)$ and $\widetilde{h}_{R,\vec{k}}(\eta)$ have to satisfy the same equation of motion, which indeed happen, since $l_R l_{\vec{k}}=l_Ll_{-\vec{k}}=1$. Thus,  equation \eqref{general_eq_for_h} for the two helicity states is not symmetric under the separate transformations $L\rightarrow R$, or $\vec{k}\rightarrow - \vec{k}$. However, it is symmetric under the simultaneous transformations:
\begin{equation}
L\rightarrow R\, \qquad {\rm and} \qquad 
\vec{k}\rightarrow -\vec{k}\,.
\label{symmetry}
\end{equation}
At this point we mention an important aspect of our analysis. The authors of \cite{Alexander:2004us} considered only the first term of the right-hand side of \eqref{eqsofmotions}, ignoring corrections that include higher than second order derivatives of the polarization tensors. In our analysis below, we shall take into account all such terms. As we shall show in the following sections of the current article, this has effects on the final estimate of the gravitational CS anomaly condensate, which during the inflationary era, considered in \cite{Alexander:2004us}, is found here to be half of the value evaluated in that work. 

We now perform the following field redefinition:
\begin{equation}
\widetilde{h}_{\lambda,\Vec{k}}(\eta) = \kappa\frac{\widetilde{\psi}_{\lambda,\Vec{k}}(\eta)}{z_{\lambda,\vec{k}}(\eta)} \,,
    \label{redefinidion}
\end{equation}
with
\begin{equation}
    z_{\lambda,\vec{k}}(\eta)=\alpha \sqrt{1 -l_{\lambda} l_{\vec{k}}  L_{CS}(\eta)}\,,
    \label{zdef}
\end{equation}
where 
\begin{equation}
    L_{CS}(\eta)=k  \xi\ , \qquad     \xi=  \frac{4 A  b^{\prime}\kappa^2}{\alpha^2}\,,
    \label{L_CS}
\end{equation}
which is dimensionless, while $[\xi] =[M]^{-1}$ represents the mass scale introduced into the system due to the existence of the gravitational anomaly term $R_{CS}$ (see \cite{Alexander:2004wk}, and references therein). Now, the perturbation tensor reads:
\begin{equation}
   h_{ij}=\kappa\sum_\lambda \int \frac{d^3k}{(2\pi)^{3/2}} e^{i\vec{k}\cdot\vec{x}}\frac{\widetilde{\psi}_{\lambda,\vec{k}}(\eta)}{\alpha \sqrt{1 - l_{\lambda} l_{\vec{k}} L_{CS}(\eta)}}\epsilon^\lambda_{ij}\,,
\end{equation}
while equation \eqref{general_eq_for_h} reduces to:
\begin{equation}
\widetilde{\psi}^{\prime\prime}_{\lambda,\Vec{k}}+ \omega_{\lambda,\vec{k}}^2(\eta)\widetilde{\psi}_{\lambda,\Vec{k}}=0,\;\;\lambda=L,R\,,
   \label{final_eq_psi}\,,
\end{equation}
with
\begin{equation}
\omega^2_{\lambda,\vec{k}}(\eta)=k^2 - \frac{z^{\prime\prime}_{\lambda,\vec{k}}(\eta)}{z_{\lambda,\vec{k}}(\eta)}
\label{omega_frequency}\,.
\end{equation}
For $\xi=0$, equation \eqref{final_eq_psi} corresponds to the wave equation of a single complex scalar field in an FLRW spacetime (in Fourier space), since $z_{\lambda,\vec{k}}=\alpha$. The presence of the CP violating coupling in \eqref{eq:Action} produces not only a time - dependent frequency for the effective harmonic oscillators, but also makes it dependent on both, the direction of propagation and helicity. The symmetry of the equations under \eqref{symmetry} is reflected on the fact that $\omega_{L,\vec{k}}=\omega_{R,-\vec{k}}$, since $z_{L,\vec{k}}=z_{R,-\vec{k}}$.\par
Using \eqref{redefinidion}, we can deduce that  $\widetilde{\psi}_{L,-\vec{k}}^*(\eta)=\widetilde{\psi}_{R,\vec{k}}(\eta)$ also holds. Thus, we can define the complex scalar field:

\begin{equation}
\begin{aligned}
&\phi(\eta,\vec{x})=\psi_L(\eta,\vec{x})=\int \frac{d^3k}{(2\pi)^{3/2}} e^{i\vec{k}\cdot\vec{x}}\widetilde{\psi}_{L,\vec{k}}(\eta),\;\; \quad \widetilde{\phi}_{\vec{k}}=\widetilde{\psi}_{L,\vec{k}}\,,\\
&\phi^*(\eta,\vec{x})=\psi_R(\eta,\vec{x})= \int \frac{d^3k}{(2\pi)^{3/2}} e^{i\vec{k}\cdot\vec{x}}\widetilde{\psi}_{R,\vec{k}}(\eta),\;\; \quad \widetilde{\phi}^*_{-{\vec{k}}}=\widetilde{\psi}_{R,\vec{k}}\,,
\end{aligned}
\label{fourierphidef}
\end{equation}
where $\phi$ and $\phi^*$ obey complex-conjugate equations of motion, as it happens in \eqref{conformaleq}. Transforming to Fourier space, the form of the complex-conjugate equation of motion translates to a difference in the equations of motion under spatial reflection of the momentum 3-vector $\vec{k}\rightarrow-\vec{k}$. 
In this sense, the corresponding action of the effective field theory for the perturbations, in Fourier space, is equivalent to that of a single complex scalar field, $\phi$, with anisotropic effective frequency, $\Omega^2_{\vec{k}}=\omega^2_{L,\vec{k}}=\omega^2_{R,-\vec{k}}\neq\Omega^2_{-\vec{k}}$,  which, of course, leads to the aforementioned property of birefringence of the perturbed CS gravity. In Fourier space, the action reads \cite{Mukhanov:2007zz}:
\begin{equation}
    S=\int d\eta \int d^3k \left(-\widetilde{\phi}^\prime_{\vec{k}}\widetilde{\phi}^{*\prime}_{\vec{k}} +\Omega^2_{\vec{k}}(\eta)\widetilde{\phi}_{\vec{k}}\widetilde{\phi}^*_{\vec{k}}\right)\,,
    \label{complexscalaraction}
\end{equation}
which indeed, as can be readily checked, produces, upon the appropriate variations, the equations of motion \eqref{final_eq_psi}, which in this formalism read
    \begin{align}
&\widetilde{\phi}_{\vec{k}}^{\prime\prime}+\Omega^2_{\vec{k}}\widetilde{\phi}_{\vec{k}}=0\label{phieom}\,,\\
&\widetilde{\phi}^{*\prime\prime}_{-\vec{k}}+\Omega^2_{-\vec{k}}\widetilde{\phi}_{-\vec{k}}^*=0\,.
\label{phistareom}
    \end{align}
We stress once again that it is the 
the gCS coupling in the action \eqref{eq:Action}, which is responsible for such anisotropic frequencies, which  
reflects its CP violating nature. On the other hand, in the absence of the gCS term, the effective descriptions of left- and right- handed polarizations in terms of scalar fields would be identical, 
of a form similar to the abovedescribed one, but with the replacement of the effective ansitropic frequency in \eqref{phieom},\eqref{phistareom} by an isotropic one, $\Omega_{\vec{k}}\rightarrow\Omega_k$, for the infinite set of complex harmonic oscillators. 

We next notice, for completion, that  the above action \eqref{complexscalaraction} can be equivalently written as:
\begin{equation}
    S=-\frac{1}{2}\sum_{\lambda=L,R} \int d\eta \int d^3k \left(  \vert \widetilde{\psi}^{\prime}_{\lambda,\vec{k}}\vert^2 - \omega^2_{\lambda,\vec{k}} \vert \widetilde{\psi}_{\lambda,\vec{k}}\vert^2 
 \right)\,,
 \label{equivalentpsiform}
\end{equation}
in agreement with \cite{Lyth:2005jf}. Finally, using \eqref{redefinidion} and \eqref{equivalentpsiform}, one can obtain the action for the gravitational-wave perturbations, up to an irrelevant total derivative, 
\begin{equation}
    S=-\frac{1}{2\kappa^2}\sum_{\lambda=L,R} \int d\eta\int d^3k \;z_{\lambda,\vec{k}}^2(\eta)\left( \vert\widetilde{h}_{\lambda,\vec{k}}^\prime\vert^2 -k^2 \vert\widetilde{h}_{\lambda,\vec{k}}\vert^2\right)\,,
    \label{GWAction}
\end{equation}
in agreement with \cite{Kamada:2019ewe}, in which the above action has been derived by keeping only up to second order terms  in the gravitational action \eqref{eq:Action}. From the above action it is clear that $L_{CS}<1$ in order to avoid ghost-like modes in the effective description \cite{Kamada:2019ewe}. The maximum value for $L_{CS}$ is acquired at the physical Ultraviolet (UV) cut-off $\mu$ of the momenta of the graviton (tensor metric perturbations) modes, which means that the latter has to be certainly bounded by the characteristic scale, $\xi$ \eqref{L_CS}, through $\mu<1/\alpha\xi$, in order to avoid pathologies in our effective description.   We stress that within our string effective actions, it is natural~\cite{Mavromatos:2022xdo} to identify $\mu$ with the string scale $M_s$. Moreover, 
consistency with the transplanckian censorship hypothesis~\cite{TCH1,TCH2,ms2}, namely that no modes exceed the Planck scale, requires that $\mu \lesssim M_{\rm Pl}$, which we assume below. 

\subsection{The $\langle R_{CS}\rangle$ Condensate of Quantum Primordial (weak) Gravitational Waves}\label{sec:cwgw}
As well known, in field theories,  (second) quantization is achieved by promoting the fields to quantum operators, and expanding them in terms of the corresponding creation and annihilation operators, which act on appropriate (Fock) state spaces. In our approach to quantum (weak) gravitational wave perturbations in this article, we have mapped the system (in the helicity basis) to 
that of a complex field \eqref{fourierphidef}.
In practice, therefore, 
we deal with the quantization of a complex scalar field in an (effectively) anisotropic background, with the action given by \eqref{complexscalaraction}. It is known that in the case of complex scalar fields, $\widetilde{\phi}^*_{-\vec{k}}\neq \widetilde{\phi}_{\vec{k}}$, with the equality holding for the case of real scalars \cite{Mukhanov:2007zz}. Thus, we are forced to introduce two sets of creation and annihilation operators, namely $\alpha^{\pm}_{\vec{k}}$ and $b^{\pm}_{\vec{k}}$, for which $(\alpha_{\vec{k}}^-)^\dagger=\alpha_{\vec{k}}^+$ and $(b_{\vec{k}}^-)^\dagger=b_{\vec{k}}^+$. These operators obey the following commutation relations:
\begin{equation}
\left[\hat{\alpha}_{\vec{k}}^-\;,\;\hat{\alpha}_{\vec{k}^\prime}^+ \right]=\left[\hat{b}_{\vec{k}}^-\;,\;\hat{b}_{\vec{k}^\prime}^+\right]=\delta^{(3)}(\vec{k}-\vec{k}^\prime)\,,
\label{creationcommutation}
\end{equation}
and zero otherwise. It is known that for the complex scalar fields $\alpha^+$ creates a particle, while $b^+$ creates its anti-particle. In this effective description the particle/anti-particle interpretation corresponds to the left and right polarizations of the gravitational waves, respectively. Furthermore, since $\phi_{\vec{k}}$ and $\phi^*_{-\vec{k}}$ obey different equations of motion, there have to be defined two sets of mode functions, which form an orthonormal basis in the complex space of solutions for each differential equation. We denote the set of mode functions for \eqref{phieom} as $\{\widetilde{v}_{\vec{k}},\widetilde{v}^{*}_{\vec{k}}\}$, while those of \eqref{phistareom} as $\{v_{\vec{k}},v^{*}_{\vec{k}}\}$, in accordance with \eqref{fourierphidef}. However, since $v^*_{-\vec{k}}$ is a solution of \eqref{phieom}, it can be written in terms of the basis $\{\widetilde{v}_{\vec{k}},\widetilde{v}^{*}_{\vec{k}}\}$. The same holds also for  $\widetilde{v}^*_{-\vec{k}}$ and $\{v_{\vec{k}},v^*_{\vec{k}}\}$. This means that $\{\widetilde{v}_{\vec{k}},v^*_{-\vec{k}}\}$ is also a basis for \eqref{phieom} and $\{v_{\vec{k}},\widetilde{v}^*_{-\vec{k}}\}$ for \eqref{phistareom}. In other words, we can just relate the bases by imposing, $\widetilde{v}_{\vec{k}}=v_{-\vec{k}}$. Then, the appropriate mode expansion reads:
\begin{equation}
\begin{aligned}
\hat{\widetilde{\phi}}_{\vec{k}}(\eta)=\widetilde{v}_{\vec{k}}\hat{\alpha}_{\vec{k}}^-+v^*_{-\vec{k}}\hat{b}_{-\vec{k}}^+\,,\\
\hat{\widetilde{\phi}}^*_{-\vec{k}}(\eta)=v_{\vec{k}}\hat{b}_{\vec{k}}^- +\widetilde{v}^*_{-\vec{k}}\hat{\alpha}_{-\vec{k}}^+\,.
\end{aligned}
\label{mode_expansion_psi}
\end{equation}
Moreover, the conjugate momenta reads:
\begin{equation}
\hat{\widetilde{\mathcal{\pi}}}_{\vec{k}}=-\hat{\widetilde{\phi}}_{\vec{k}}^{*\prime} \,,
\end{equation}
obeying the commutation relation
\begin{equation}
    \left[ \hat{\widetilde{\phi}}_{\vec{k}}(\eta), \hat{\widetilde{\mathcal{\pi}}}_{\vec{k}^\prime}(\eta)  \right]=i\delta(\vec{k}-\vec{k}^\prime)
    \label{phiandmomentum}
\end{equation}
Then, in order to be consistent with the commutation relations \eqref{creationcommutation}, \eqref{phiandmomentum}, the mode functions should obey the following Wronskian normalization condition:
\begin{align}
    & \widetilde{v}_{\vec{k}} \widetilde{v}^{*\prime}_{\vec{k}} - \widetilde{v}^*_{\vec{k}} \widetilde{v}^{\prime}_{\vec{k}} =-i\label{Wronkian1}\,,\\
    &v_{\vec{k}} v^{*\prime}_{\vec{k}} - v^*_{\vec{k}}v^{\prime}_{\vec{k}} =-i\,.
\end{align}
We now remark that the gravitational tensor perturbations corresponding to 
the gravitational waves are also quantized, within the weak quantum gravity context, according to \eqref{redefinidion} and \eqref{fourierphidef}:
\begin{equation}
    \begin{aligned}
        & \widetilde{h}_{L,\vec{k}}= u_{L,\vec{k}}\alpha^-_{\vec{k}}+u^*_{R,-\vec{k}}b^+_{-\vec{k}}\,,\\
        & \widetilde{h}_{R,\vec{k}}= u_{R,\vec{k}}b^-_{\vec{k}}+u^*_{L,-\vec{k}}\alpha^+_{-\vec{k}} \,,
    \end{aligned}
    \label{hL_hR}
\end{equation}
where 
\begin{equation}
\begin{aligned}
&u_{L,\vec{k}}=\kappa\frac{\widetilde{v}_{\vec{k}}}{z_{L,\vec{k}}}\,,\\
& u_{R,\vec{k}}=\kappa\frac{v_{\vec{k}}}{z_{R,\vec{k}}}
\label{modefinction_for_h}\,.
\end{aligned}
\end{equation} 
Then, correlators between the helicity basis are non-zero, having the following form:
\begin{align}
    \langle \widetilde{h}_{R,\vec{k_1}}\widetilde{h}_{L,\vec{k_2}}\rangle=u_{R,\vec{k_1}}u^*_{R,-\vec{k_2}}\delta(\vec{k_1}+\vec{k_2})
    \label{correlator1}\,,\\
    \langle \widetilde{h}_{L,\vec{k_1}}\widetilde{h}_{R,\vec{k_2}}\rangle=u_{L,\vec{k_1}}u^*_{L,-\vec{k_2}}\delta(\vec{k_1}+\vec{k_2})\,.
    \label{correlator2}
\end{align}

The condensate of the gravitational anomaly can be computed according to the above considerations as follows. In conformal time, we have: 
\begin{equation}
   \langle R_{CS}\rangle= \frac{2i}{\alpha^4}\left[ \langle\partial^{2}_z h_L\partial_z h_{R}^{\prime}\rangle+ \langle h^{\prime\prime}_L\partial_z h^{\prime}_R\rangle -\langle\partial^{2}_z h_R\partial_z h^{\prime}_L\rangle- \langle h^{\prime\prime}_R\partial_z h^{\prime}_L \rangle\right]\,,
    \label{RCSconformal}
\end{equation}
where the prime denotes differentiation with respect to the conformal time. Having as a rule in the Fourier space that derivatives with respect to $z$ mean multiplication with $ik$, while derivatives with respect to the conformal time mean derivatives of the mode functions, one can calculate each term having as a guide the equations \eqref{correlator1} and \eqref{correlator2}. With these in mind, we then obtain, 
\begin{align}
&\left<\partial^{2}_z h_R(\Vec{x},\eta)\partial_z h^{\prime}_L(\Vec{x},\eta)\right>=-\int^{\alpha\mu}\frac{d^3 \Vec{k}}{(2\pi)^3}l_{\vec{k}}(i k^3)  u_{R,\vec{k}}  u^{*\prime}_{R,\vec{k}} \,,\\ 
&\left<h^{\prime\prime}_R\partial_z h^{\prime}_L \right>  = \int^{\alpha\mu}\frac{d^3 \Vec{k}}{(2\pi)^3}l_{\vec{k}}(i k)  u^{\prime\prime}_{R,\vec{k}}u^{*\prime}_{R,\vec{k}} \,,\\
&\left<\partial^{2}_z h_L(\Vec{x},\eta)\partial_z h^{\prime}_R(\Vec{x},\eta)\right>=-\int^{\alpha\mu}\frac{d^3 \Vec{k}}{(2\pi)^3}l_{\vec{k}}(i k^3)  u_{L,\vec{k}}  u^{*\prime}_{L,\vec{k}} \,,\\ 
&\left<h^{\prime\prime}_L\partial_z h^{\prime}_R \right>  =  \int^{\alpha\mu}\frac{d^3 \Vec{k}}{(2\pi)^3}l_{\vec{k}}(i k)  u^{\prime\prime}_{L,\vec{k}}u^{*\prime}_{L,\vec{k}}\,.
\end{align}
Substituting to \eqref{RCSconformal}, we obtain:
\begin{equation}
     \langle R_{CS}\rangle = \frac{2}{\alpha^4}\int^{\alpha\mu}\frac{d^3 \Vec{k}}{(2\pi)^3}l_{\vec{k}}\left[ k^3  \left(u_{L,\vec{k}}  u^{*\prime}_{L,\vec{k}}-u_{R,\vec{k}}  u^{*\prime}_{R,\vec{k}} \right) + k \left(u^{\prime\prime}_{R,\vec{k}}u^{*\prime}_{R,\vec{k}}-u^{\prime\prime}_{L,\vec{k}}u^{*\prime}_{L,\vec{k}}\right)\right]\,.\\
     \label{Rcs_condensate}
\end{equation}
Since each polarization satisfies a different equation, the above condensate is non-vanishing.

\subsubsection*{The $\langle R_{CS}\rangle$ Condensate in a Stiff Background}\label{stiff_era_condensate}

Now, we can choose specifically the era of interest by specifying the dependence of the scale factor with respect to the conformal time, and also by choosing our background through the equations of motion for the field $b$. 
The above developed formalism therefore allows us to discuss in detail the passage from the stiff KR axion era to the RVM inflationary era,
within the StRVM cosmology, which had only been sketched in \cite{bms,ms1,ms2}.
Specifically, for the Stiff Era, the background equations of motion, read: 
\begin{align}
    &3\left(\frac{\dot{\alpha}}{\alpha}\right)^2=\frac{\kappa^2}{2}\dot{b}^2\label{grav_back1}\,,\\
&\left(\frac{\dot{\alpha}}{\alpha}\right)^2+2\frac{\ddot{\alpha}}{\alpha}=-\frac{\kappa^2}{2}\dot{b}^2\label{grav_back2}\,,\\
   &\ddot{b}+3\frac{\dot{\alpha}}{\alpha}\dot{b}=0
   \,.\label{axion_back}
\end{align}
 The solution for the field and the scale factor dependence on the conformal time $\eta$ are given by: 
\begin{align}
    & b(\alpha)=  \frac{\sqrt{6}}{\kappa}\log\left(\alpha\right) \label{axion_stiff_era} + b(\eta_0)\,,\\ 
    & \alpha(\eta)=\sqrt{\eta/\eta_0} 
    \label{scale_factor_stiff_era}\,,
\end{align}
respectively, 
where the dots denote derivatives with respect the FLRW cosmic time, $t$, and  $\eta_0$ is a constant, which represents the beginning of the stiff era in the StRVM cosmology, at which we assume the boundary condition $b(\eta_0) <0$.
The negative value of the initial condition of the field $b(\eta_0)$
at the beginning of the stiff era, 
is necessitated by the requirement that  $b(\eta_i)$, where $\eta_i \gg \eta_0$ denotes the end of the stiff era and the onset of the RVM inflation, must be negative, while 
$\dot b(t) >0$ throughout, as discussed in section \ref{sec:EN} (see~Figure \ref{b_and_bdot_infl_fig}). Such results are also in  agreement with the considerations of \cite{bms,ms1,ms2,Mavromatos:2022xdo}.

The actual value of $\eta_0$ cannot be determined for a generic stiff era, as its duration actually depends on the details of the microscopic string model that underlies the phenomenological StRVM cosmology of \cite{bms,ms1,ms2}.
If we assume a first hill-top inflation, for instance, as done in \cite{ms1}, then $\eta_0$ represents the exit from that phase, at which the massless KR axion dominance is assumed to take place, with these excitations being created by the decay of the vacuum during the first inflation. The details of the latter are not important for our purposes in this work. 
The end of the stiff era, during which the condensate of the gCS term $\langle R_{CS}\rangle $ forms, can be consistently defined by $\eta \gg \eta_0$, without specifying further the magnitude of $\eta_0$.

Now, we want to estimate the order of the dimensionless quantity $L_{CS}(\eta)$ appearing in eq. \eqref{L_CS}, considering the physical UV cut off $\mu$ of our theory to be the string scale, $M_s$, {\it i.e}. 
modes with mometa $k \sim \alpha M_s =\mu$. Keeping in mind that in the StRVM the CS coupling $A$ ({\it cf.} \eqref{Aval}) is of order,
\begin{equation}
      A \sim 10^{-2} \ \mathcal{O}\left(\frac{M_{\rm Pl}}{M_{s}^2}\right)\,,
\end{equation}
the maximum order attained by $L_{CS}$ is:
\begin{align}
    L_{CS} \sim 10^{-2}\left(\frac{\dot{b}}{M_s M_{\rm Pl}}\right)\,,
    \label{maxLCS}
\end{align}
which is realised for modes with momenta near the UV cut-off $\mu$. We must assume the following condition $L_{CS} << 1$. Since, $b^\prime\sim 1/\eta$, the statement of $L_{CS} << 1$  can be understood as a late-stiff era condition, i.e. the condensate forms after sufficient time of evolution according to the stiff equation of state, $w_b=+1$, during the pre-RVM inflationary era of the StRVM~\cite{ms1,ms2}.
Keeping now only the leading order term, with respect to $L_{CS}$, in eq.~\eqref{final_eq_psi} during the stiff era, we obtain: 
\begin{equation}
    \frac{d^2}{dx^2}\psi_{\lambda,\Vec{k}} + \left( 1+\frac{1}{4x^2}\left[1 +8 l_\lambda l_{\vec{k}} L_{CS} +\mathcal{O}\left(  L_{CS}^{2}\right)\right]\right)\psi_{\lambda,\Vec{k}}=0\,,
\end{equation}
where $x \equiv k\eta$. Thus, since $L_{CS}<<1$, it suffices to solve the following equation for the mode functions of $\psi_\lambda$,
\begin{equation}
\frac{d^2}{dx^2}\psi_{\lambda,\Vec{k}} + \left( 1+\frac{1}{4x^2}\right)\psi_{\lambda,\Vec{k}}=0
    \label{wave_eq_psi}\,,
\end{equation}
with the final contribution of the gravitational anomaly introduced through the mode functions \eqref{redefinidion}. In this case, we have abandoned the dependence of the equation of the helicity, something that is valid for the late stiff era approximation we are concerned with. This means that we have only one differential equation, the same for $\psi_{L,\Vec{k}}$ and $\psi_{R,\Vec{k}}$. Consequently, the two sets of mode functions defined for \eqref{phieom} and \eqref{phistareom} degenerate into one common basis, i.e. $\{v_{k},v^{*}_{k}\}$, with the Wronskian normalization:
\begin{equation}
    v_{k} v^{*\prime}_{k} - v^*_{k}v^{\prime}_{k} =-i\,.
    \label{WronskianIsotropic}
\end{equation} 
The solution of \eqref{wave_eq_psi}, reads:
\begin{equation}
    v_{k}(\eta)= \sqrt{ \eta}\left[C_1
    J_0(k\eta)+C_2 Y_0(k\eta)\right]
    \label{sol_bessel_mu}\,,
\end{equation}
with $C_1,C_2$  dimensionless integration constants and $J_0, Y_0$ are the Bessel functions of the first and second kind~\cite{Gradshteyn:1943cpj}. As $\eta \to \infty$, the effective frequency $\omega^2_{\lambda,\vec{k}}(\eta)\rightarrow k^2$ and thus the solution should have a smooth connection with the Minkowski mode function (Bunch Davies vacuum) that obeys the Wronskian normalization \eqref{WronskianIsotropic}  \cite{Mukhanov:2007zz} 
\begin{equation}
v_{k}(\eta)|_{\eta \to \infty}=\frac{1}{\sqrt{2 k}}e^{i k \eta}\,.
\label{MinkowskiMatch}
\end{equation}
The asymptotic behavior of the Bessel functions of equation \eqref{sol_bessel_mu} is given by:
\begin{align}
    &J_0 (k \eta) \to \sqrt{\frac{2}{\pi}}\sqrt{\frac{1}{k \eta}}\cos\left(k\eta-\frac{\pi}{4}\right)\,, \\ 
    &Y_0 (k \eta) \to \sqrt{\frac{2}{\pi}}\sqrt{\frac{1}{k \eta}}\sin\left(k\eta-\frac{\pi}{4}\right)\,.
\end{align}
So, the asymptotic limit of $v_{\lambda,k}$ reads:
\begin{equation}
    v_{k}(\eta)|_{\eta \to \infty}\to \frac{1}{\sqrt{\pi k}}\left[ (C_1-C_2) \cos k\eta + (C_1+C_2)\sin k\eta   \right]\,.
\end{equation}
Matching with \eqref{MinkowskiMatch}, requires that $C_1 =\sqrt{\pi} e^{i\pi/4}/2$ and $C_2 = i \ C_1$. 
We finally obtain then:
\begin{align}
    &v_{k}(\eta)= e^{i\pi/4} \frac{\sqrt{\pi \eta}}{2}\left[J_0(k\eta)+i Y_0(k\eta)\right]
    \label{sol_bessel_2}\,, \\
    &v^{*}_{k}
    (\eta)=e^{-i\pi/4} \frac{\sqrt{\pi \eta}}{2}\left[J_0(k\eta)- i Y_0(k\eta)\right]
    \label{sol_bessel_2*}\,.
\end{align}
The redefinition functions $z_{\lambda,k}(\eta)$ \eqref{zdef} during the stiff era are given by:
\begin{align}
    z^{stiff}_{L,\vec{k}}(\eta)=\sqrt{\frac{\eta}{\eta _0}+l_{\vec{k}} \frac{2 \sqrt{6} A \kappa  k}{\eta}}\,, \\ 
    z^{stiff}_{R,\vec{k}}(\eta)=\sqrt{\frac{\eta}{\eta _0}-l_{\vec{k}} \frac{2 \sqrt{6} A \kappa  k}{\eta}}\,,
\end{align}
and, so, from the field redefinition \eqref{redefinidion} and eq.~\eqref{modefinction_for_h}, we obtain the mode functions $u_{L,\vec{k}},u_{R,\vec{k}}$ for the gravitational waves:
\begin{equation}
    \begin{aligned}
    &u_{L,\vec{k}}=\kappa\frac{e^{i\pi/4}}{2}\frac{\sqrt{\pi\eta}}{\sqrt{\frac{\eta}{\eta _0} + l_{\vec{k}} \frac{2 \sqrt{6} A \kappa  k}{\eta}}}\left[J_0(k\eta)+i Y_0(k\eta)\right]\,, \\
    &u_{R,\vec{k}}=\kappa\frac{e^{i\pi/4}}{2}\frac{\sqrt{\pi\eta}}{\sqrt{\frac{\eta}{\eta _0} - l_{\vec{k}}\frac{2 \sqrt{6} A \kappa  k}{\eta}}}\left[J_0(k\eta)+i Y_0(k\eta)\right] \,.
    \label{modefunctions_h}
\end{aligned}
\end{equation}

We can calculate now the gravitational condensate of eq.~\eqref{Rcs_condensate} during the late stiff era that precedes RVM inflation in the StRVM model of \cite{bms,ms1,ms2}. To this end,  we shall only keep first order terms with respect to  $L_{CS}$. Furthermore, since we are interested to the late stiff era $\left(\eta \gg \eta_0\right)$, we will use the asymptotic behavior for the mode functions in order to calculate the integral of eq. \eqref{Rcs_condensate}. With the physical UV cut-off given by $k \approx \alpha \mu$, and applying the aforementioned approximations to eq.~\eqref{Rcs_condensate}, we obtain the final result:
\begin{equation}
      \left<R_{CS}\right>^
      {\rm stiff}=-\frac{1}{2}\langle  R_{\mu\nu\rho\sigma}\, \widetilde R^{\mu\nu\rho\sigma}\rangle=-\frac{30\sqrt{6} A \kappa ^3 \mu ^4}{\pi ^2} H_{\rm stiff}(\eta)^4
      \label{VEV_stiff}\,.
\end{equation}
In the above expression, $H(\eta)$ corresponds to the Hubble rate of a stiff-matter dominance era, as a function of the conformal cosmic time, which reads:
\begin{equation}\label{Hshift}
  H_
  {\rm stiff} (\eta)=\frac{\alpha^{\prime}}{\alpha^2}=\frac{\sqrt{\eta_0}}{2  \eta^{3/2}}\,.
\end{equation}
In arriving at \eqref{VEV_stiff}, we have tacitly assumed that the number density $\mathcal N_S$ of sources that produce this condensarte is of $\mathcal{O}(1)$. In \cite{Mavromatos:2022xdo}, it has been suggested that in general the number of sources is a phenomenological parameter, which actually depends on the microscopic string model underlying StRVM~\cite{ms1,ms2}, and in fact, a non trivial density is necessary in order to be able to consistently implement a UV cutoff for the modes $\mu \simeq M_s$, without restricting the string scale $M_s$, as required by viewing the StRVM as an effective low energy gravitational theory stemming from strings.   
With this in mind, the gCS condensate \eqref{VEV_stiff} during the stiff era has to be replaced by the more general expression:
\begin{equation}
\left<R_{CS}\right>^{total}_
      {\rm stiff}= -\mathcal N_S \, \frac{30\sqrt{6} A \kappa ^3 \mu ^4}{\pi ^2} H_{\rm stiff}(\eta)^4
      \label{VEV_stiff2}\,,
\end{equation}
assuming, as in \cite{Mavromatos:2022xdo}, a linear superposition of the effects of the various sources when evaluating the condensate as a vacuum expectation value stemming from  weak quantum graviton modes (tensor perturbations of chiral gravitational wave type).

\section{Inflation induced from gravitational condensates - A Dynamical System approach}\label{sec:infl} 

Based on the previous analysis, we shall present in this section a detailed demonstration as to how the condensation of the gravitational waves, assumed to occur at the end of the stiff era of the StRVM, as discussed in section \ref{sec:cwgw} above,  can lead to a transition from the stiff era to an inflationary universe (of RVM type). After the condensation of the chiral gravitational waves, the gravitational effective action will be given by \eqref{eq:Action}, with the condensate given by \eqref{VEV_stiff}. In this sense, a linear effective potential for the axion arises~\cite{Mavromatos:2022xdo,Mavromatos:multiaxion}:
\begin{equation}
    V_{eff}=  A\langle R_{CS}\rangle b\,,
    \label{effectivepot}
\end{equation}
which breaks the shift symmetry \eqref{shiftsym} of the Lagrangian \eqref{sea3}. Such a potential is identical with the one in \eqref{linearpotential}, where the corresponding constant is given by $\mathcal{C}=A\langle R_{CS}\rangle$. During the stiff era, i.e. in the absence of gravitational waves, the gCS term is identically zero for an FLRW background. This is the case where the solution according to the EN variables, described in section \ref{sec:EN},
is $x=\frac{\kappa \dot{b}}{\sqrt{6}H}=1$, and  $y=0$. Before the introduction of the linear effective potential, the system lies on one of the two critical points $O(0,0)$ or $C(\pi,0)$, with an equation of state $\omega_b=+1$. In other words, the state of the KR-axion cosmic fluid before the influence of the potential, is either at the left or at the right circles in the abscissa of Figure~\ref{stream}. As we have to deal with unstable fixed points, stiff solutions for our system are like a pencil balancing on its tip. A slight perturbation on it could drive the system to another phase. Such a perturbation is introduced by the presence, and eventual condensation, of the chiral gravitational waves \eqref{perturbationmatrix}.

In order to determine the order of magnitude of this perturbation, we first express the order of magnitude of the square of our EN variable $y$, which is given by \eqref{ENvariables}, but now using the effective potential  \eqref{effectivepot}: 
\begin{equation}
    y_i^2 \equiv {\rm sin}^{2}\varphi_i = \frac{\kappa^2\vert V_{eff}\vert}{3H_{i}^2}
\end{equation}
where $H_{i}$ is the Hubble rate at the formation of the condensate. Thus, the order of magnitude for $y^2$, can be expressed, through \eqref{VEV_stiff}, as 
\begin{equation}
\mathcal{O}\left(y_i^2\right)
\equiv \mathcal{O}({\rm sin}^{2}\varphi_i) \sim   2.5 \times 10^{-4}\left(\frac{H_{i}}{M_{\rm Pl}}\right)^{2}\left(\frac{\abs{b_i}}{M_{\rm Pl}}\right)\,,
\label{orderofmagnitude}
\end{equation}
where, as already mentioned, we considered that the UV  cutoff of our theory is provided by the string scale $M_s$, $\mu \approx M_s$~\cite{Mavromatos:2022xdo}, which is consistent with viewing StRVM as a low-energy string effective gravitational theory. Assuming an order of magnitude of $\varphi_i\sim 10^{-p}$, $p \in \mathbb R$, one can easily ascertain, following the analysis in section \ref{linear_potential}, that, for $\zeta_i\sim10^{-2}$, the Hubble rate at inflation is related to its initial value $H_i$, at the moment of the formation of the CS condensation in the shift era,  through $H_i\sim 10^{p+1}H_I$. Moreover, in view of \eqref{zetalinear}, $\zeta_i\sim10^{-2}$ implies:
\begin{equation}\label{b0}
    \frac{\abs{b_i}}{M_{\rm Pl}}\sim \mathcal{O}\left(10\right)\,.
\end{equation}
The upper bound on the Hubble rate at inflation, imposed by the Planck Collaboration data \cite{Planck:2018vyg}, sets an order of magnitude $H_I \sim 10^{-5}M_P$. On account of  \eqref{orderofmagnitude}, this yields 
\begin{equation}
   \varphi_i\approx  10^{-5/2} \,.
\end{equation} 
In such a case, the gravitational condensate has to be formed at the stiff era in a phase where the Hubble rate is about 
\begin{equation}
    H_{i} \sim 10^{7/2}\, H_I \approx10^{-3/2}M_{\rm Pl} \,.
\end{equation}
Furthermore, from eq.\eqref{axion_stiff_era}, we obtain that, at the formation of the condensate, one has:
\begin{equation}\label{bdot}
    \dot{b}_i= M_{\rm Pl}\sqrt{6}H_i\sim 7.7 \times 10^{-2} M^2_{\rm Pl} \sim  10^{-1} M^2_{\rm Pl}\,,
\end{equation}
implying from eq.\eqref{maxLCS} that $ L_{CS}\sim 7.7\times 10^{-4}M_{\rm Pl}/M_s $. Thus, by 
assuming the maximum order of  $L_{CS}\sim 10^{-2}$ \eqref{maxLCS}, as necessary for the formation of the condensate at the late stiff era, we obtain for the string scale (cut-off) the order of   
\begin{equation}
    M_s\sim 10^{-1}M_{\rm Pl}<M_{\rm Pl}\,,
    \label{cutoff}
\end{equation}
consistent with the transplanckian transplanckian censorship hypothesis~\cite{TCH1,TCH2,ms2}.

\subsection{The constancy of $\langle R_{CS}\rangle$}
We can now apply the same argumentation in order to calculate the gravitational condensate during inflation, in order to study how this condensate can remain constant. 
The redefinition function is given by \eqref{zdef} and \eqref{L_CS},
with the scale factor given by $\alpha(\eta)=-1/H_I \eta$, where $\eta<0$ is the conformal time during inflation. Writing again eq.\eqref{final_eq_psi} up to leading order to the small, dimensionless quantity $L_{CS}$, we have:
\begin{equation}
    \frac{d^2}{dx^2}\widetilde{\psi}_{\lambda,\Vec{k}}+\left[1-\frac{2}{x^2}\left(1-l_\lambda l_{\Vec{k}}\frac{L_{CS}}{2}+\mathcal{O}\left(L_{CS}^2\right)\right)\right]\widetilde{\psi}_{\lambda,\Vec{k}}=0
\label{psi_inflation_dimensionless}
\end{equation}
where $x=k\eta$. We can see that, during inflation, $\abs{L_{CS}}=\widetilde{L}_{CS}\cdot \vert x\vert$, with $\widetilde{L}_{CS}=4 A \dot{b} \kappa^2 H_I\sim 10^{-11}$. We can check that around the cut-off of the effective theory defined from \eqref{cutoff}, the maximum value for $x = k \abs{\eta} \sim 10^{4}$.  This gives the maximum value of $L_{CS}$:
\begin{equation}
    L^{max}_{CS}\sim 10^{-7}<<1
\end{equation}
Thus, equation \eqref{psi_inflation_dimensionless} can be approximated by:
\begin{equation}
    \frac{d^2}{dx^2}\widetilde{\psi}_{\lambda,\Vec{k}}+\left(1-\frac{2}{x^2}-l_\lambda \frac{\widetilde{L}_{CS}}{x}\right)\widetilde{\psi}_{\lambda,\Vec{k}}=0
    \label{inflation_wave equation}
\end{equation}
The terms of inverse powers of $x$ are important only for the super - horizon modes, $x<<1$. Considering only the sub-horizon modes, as in \cite{Alexander:2004us}, the solution  of eq.\eqref{inflation_wave equation}  can be approximated by the plane waves:
\begin{align}
     &v_k=\frac{1}{\sqrt{2 k}} e^{ik\eta} \,, \\ 
     &v^{*}_{k}=\frac{1}{\sqrt{2 k}}e^{-ik\eta}\,,
    \label{sol_inflation}
\end{align}
that is, again the equation is reduced to the isotropic case, while the anisotropic contribution is introduced through \eqref{modefinction_for_h}, with the two bases reduced again to the isotropic one, $\{v_k,v^*_k\}$, which solves \eqref{inflation_wave equation}. The redefinition function of eq. \eqref{zdef} during inflation is given by:
\begin{equation}
   \begin{aligned}
    &z^{I}_{L,\vec{k}}(\eta)=-\frac{1}{H_I\eta}\sqrt{1-l_{\vec{k}}4 A \dot{b} H_I ^2 k \kappa^2\eta}\,, \\ 
    &z^{I}_{R,\vec{k}}(\eta)=-\frac{1}{H_I\eta}\sqrt{1+l_{\vec{k}}4 A \dot{b} H_I ^2 k \kappa^2\eta}\,.
   \end{aligned} 
\end{equation}
Using  \eqref{modefinction_for_h}, we can calculate the gCS condensate during inflation. On expanding the quantity inside the integral of eq. \eqref{Rcs_condensate} up to leading order for $L^{I}_{CS}$ and performing the integration, we obtain the result:
\begin{equation}
      \langle R_{CS} \rangle^I =-\frac{ A }{\pi ^2}\frac{\dot{b}_I}{M_{\rm Pl}} \left(\frac{H_{I}}{M_{\rm Pl}}\right)^3 \mu ^4<0\,,
      \label{VEV_inflation}
\end{equation}
 where the negative sign guarantees a positive sign for the  effective cosmological constant \eqref{effective_cosmological_constant}.  It is important to note that the above result differs by a factor of two from the one presented
in \cite{bms,ms1,ms2,Mavromatos:2022xdo,Mavromatos:multiaxion}, which is based on the result of \cite{Alexander:2004us}. The reason for the discrepancy lies on eq.\eqref{eqsofmotions}. As already mentioned in section \ref{sec:gwcs}, the authors of \cite{Alexander:2004us} considered only the first term of the right-hand side of eq.\eqref{eqsofmotions} and ignored corrections that include higher order spatial derivatives. In our analysis, we have taken into account every contribution of the equations of motion for the gravitational waves, even those coming from higher order derivatives. 
If one considers only up to first order spatial derivatives to \eqref{eqsofmotions}, the corresponding redefinition function has the form $z=a(\eta)e^{\pm 4 A\dot{b}k\kappa^2/a(\eta)}$. Consequently, the correction due to the CS coupling, up to first order to $L_{CS}$, reads, $z\approx \alpha(\eta)\pm 4 A\dot{b}k\kappa^2 $. Such a result yields a contribution which is two times larger than that of the redefinition function \eqref{redefinidion}, leading to a two times larger result for the condensate \eqref{VEV_inflation}, which was the case of \cite{Alexander:2004us}.  Nonetheless, the order of magnitude of the condensate remains the same, and hence the phenomenology of the StRVM~\cite{bms,ms1,ms2,Mavromatos:2022xdo,Mavromatos:multiaxion}, is not qualitatively affected.

As in the case of the stiff matter era \eqref{VEV_stiff2}, for the actual value of the condensate in the inflationary epoch, we have to multiply the result \eqref{VEV_inflation}
with the proper number density $\mathcal{N}_I$ of all kinds of sources of gravitational waves during inflation, obtaining as a final result~\cite{Mavromatos:2022xdo}:
\begin{equation}
      \left<R_{CS}\right>^{total}_{I}=-\mathcal{N}_I\frac{ A  \kappa ^4 \mu ^4}{\pi ^2}\dot{b}_I H_{I}^3\,.
      \label{VEV_inflation_source}
\end{equation}

 As already mentioned, eq.~\eqref{b_bdot_orders} (see also Figure \ref{b_and_bdot_infl_fig}), 
 provides an estimate of the order of magnitude of $\dot{b}$, which remains approximately constant during the entire inflation era: 
\begin{equation}
    \dot{b}_I\sim 10^{-1} H_I M_{\rm Pl}\,,
    \label{bdotInflation}
\end{equation}
confirming in this way the assumption made in \cite{bms,ms1,ms2,Mavromatos:2022xdo}. This is a highly non-trivial consistency check of the StRVM approach to inflation via primordial-gravitational-wave-induced CS condensates, which in this way is mapped into a dynamical evolution of a single-(axion) field system with a linear potential.\footnote{We mention for completeness that our calculations in this work are also in agreement with the requirement that the axion is of order $\abs{b_I}\sim\mathcal{O}(10)M_{\rm Pl}$ and does not change order of magnitude during inflation, which is a crucial fact for the string-inspired Lorentz- and CPT- Violating Standard Model Extension~\cite{kostel}, which stems as a consequence of the StRVM framework~\cite{bms,ms1,ms2,Mavromatos:2022xdo}.}\color{black}

Moreover, assuming the constancy of the value of the condensate during the entirety of the inflationary era, and using eq.\eqref{bdotInflation}, 
we have to match eq.\eqref{VEV_inflation_source} with eq.\eqref{VEV_stiff2}. Then, we obtain: 
\begin{equation}
   \frac{\mathcal{N}_I}{\mathcal{N}_S}\sim 7\cdot 10^2 \left(\frac{H_i}{H_I}\right)^4\,,
\end{equation}
where, we recall, $\mathcal{N}_S$ denotes the proper number density  of sources of gravitational waves during the stiff era, that form the condensate. In the result of eq. \eqref{VEV_stiff}, we assumed $\mathcal{N}_S \sim \mathcal{O}(1)$, but we could in general consider an enhancement in this period, i.e. from populations of primordial black holes. Such an enhancement would have impact on our theory, since it would affected the time of creation of the gravitational condensate and the upcoming inflationary evolution, but it would also influence the cutoff scale of our theory \eqref{cutoff}. For the initial condition we have already assumed, for which $H_i/H_I\sim 10^{7/2}$, we obtain,
\begin{equation}
    \frac{\mathcal{N}_I}{\mathcal{N}_S}\sim 7\cdot 10^{16}\,,
\end{equation}
which lies in the range given in \cite{Mavromatos:2022xdo}, stemming from the assumption of the constancy of the CS condensate during inflation, which thus avoids exponential dilution~\cite{bms,ms1}.

Before closing this section, we would like to stress once more that the presence of the UV cutoff $\mu$, which is identified with the string scale $M_s$ in our approach, is unavoidable, because our gravitational theory is viewed as an effective field theory obtained from strings, which is valid at energy scales up to $M_s$. In this respect, our model, and that of \cite{bms,bms2,ms1,ms2}, is different from purely local effective field theory models of gravitational leptogenesis, 
studied in \cite{Fischler:2007tj,Kamada:2020jaf},
where the gCS condensate  can be estimated by adding to the theory appropriately (and not uniquely) chosen counterterms (in specific expanding universe backgrounds) to cancel the divergent terms that are proportional to the UV cutoff scale. Such a procedure~\cite{Kamada:2020jaf} leaves a finite value of the condensate, which, in contrast to our string theory case \eqref{VEV_stiff}, \eqref{VEV_inflation_source}, \eqref{bdotInflation}, is found to depend on the eighth power of the Hubble parameter. Par contrast, in  our case the condensate is proportional to $H^4$.  
In our string theory context, adding such local effective counterterms to remove the UV cutoff, has no meaning, as we have already mentioned.  The presence of the UV cutoff in our case, which is identified with the string scale $M_s$, signals the r\^ole of infinite towers of massive string states (hence non local, purely stringy effects). 
As discussed in \cite{Mavromatos:2024pho}, from the point of view of a low energy observer, such towers can be viewed  as an ``environment'', which is reflected in the presence of imaginary parts in the condensate. The latter leads to an instability of the corresponding de Sitter vacuum, thus allowing exit from inflation, consistent with the dynamical system approach. The imaginary parts can lead to 
an estimate of the lifetime of inflation, consistent with phenomenology.

\section{Conclusions and Outlook}\label{sec:concl} 

In this work we have revisited the string-inspired cosmological Chern-Simons (CS) gravitational theory  
that corresponds to the Stringy Running Vacuum Model (StRVM) Cosmology, proposed in \cite{bms,ms1,ms2}. We have re-evaluated the gravitational anomaly condensate induced  by primordial gravitational waves (PGWs), 
by relaxing some of the approximations that have been employed in \cite{Alexander:2004us,Lyth:2005jf}. As we have shown, there is considerable reduction (by a half) of the value of the condensate, as compared with the approximate evaluation of \cite{Alexander:2004us,Lyth:2005jf}, but since the order of magnitude remains the same, the conclusions on the cosmology of the model in the analysis of \cite{bms,ms1,ms2} are not affected. 

In our analysis, we have used dynamical systems to study both the KR-axion-stiff and  RVM inflationary phases of the model, and the transition from the former to the latter in detail, which has not been done before. 
The formation of the CS anomaly condensate leads to an effective gravitational theory with a linear KR axion potential, whose dynamics is studied using the aforementioned dynamical system approach.  
In this context, we should mention, that to be complete, one should estimate the order of magnitude of the imaginary parts of the quantum effective action, obtained from integrating our graviton degrees of freedom~\cite{Lyth:2005jf,ms2}. This would in principle yield an estimate of the life time of the metastable inflationary era,\footnote{It is worthy of mentioning that the metastability of the inflationary era is a welcome fact from the point of view of the StRVM, as implying compatibility, at least in principle, with the swampland criteria~\cite{swamp1,swamp2,swamp3,swamp4,swamp5,srvmswamp} for embedding the theory into a consistent quantum gravity framework.} which, for phenomenological consistency of the model, should produce a number of e-foldings $N  \gtrsim 50-60$. There are subtleties in such a computation in the context of our weak quantum gravity models,
especially those associated with gauge invariance, which are additional to the already technical complications of flat space-time models~\cite{decay_1,decay_2,decay_3,decay_4,decay_5,decay_6,decay_7}. This is left for future work, noting that a rough estimation of the associated life time for RVM inflation in our model has been given in \cite{Mavromatos:2024pho}.

There are several interesting avenues of research that we would like to pursue, related to issues which, although briefly outlined in \cite{bms,ms1,ms2}, nonetheless  have never been studied in detail. Among the most important of them is the existence of linear in cosmic time KR axion backgrounds, which are responsible for a spontaneous breaking of (local) spacetime Lorentz invariance during the presence of the primordial-gravitational-wave-induced CS condensate. 
As argued in \cite{bms,ms1}, such backgrounds remain undiluted during the RVM inflation, surviving intact during the radiation era that succeeds the inflationary epoch. In models with massive right-handed neutrinos in their matter sector, then, the presence of these linear KR axion backgrounds leads to the creation of a Lepton asymmetry (Leptogenesis), according to the mechanism suggested in \cite{deCesare:2014dga,Bossingham:2017gtm,Bossingham:2018ivs,Mavromatos:2020dkh}. Eventually this leads to baryogenesis, via, say, Baryon(B)- and Lepton(L)- number violating, but B-L conserving, sphaleron processes in the standard model sector of the effective matter theory. Although for exactly constant CS condendsates, such KR axion backgrounds exhibit a 
constant rate, nonetheless in actual situations, as a result of the cosmic-time dependence of the CS condensate 
 during the end of the RVM inflationary period, these KR axion fields are characterised by time dependent rates.
It would be interesting to exploit the precise effects of such time dependence of the KR axion background on the leptogenesis processes. 

Another potentially interesting issue, is an estimation of the 
duration of the transition of the system from the end of the RVM inflationary era to the radiation era. This depends  
on details of the microscopic string theory that underlies the StRVM.
In general, depending on their parameters, RVM cosmologies may be characterised by long reheating phases~\cite{Lima:2013dmf,Lima:2015mca}. In the case of string theory, there are additional reasons which might prolong the exit phase from inflation, as the latter depends
on the detailed axion dynamics within the StRVM framework. Specifically, given that StRVM is a string-inspired theory, its actual axion spectrum is much more complicated than the single KR axion model we discussed above and in \cite{bms,ms1,ms2}. Indeed, in string theory, in addition to the string-model independent KR axion, one faces a multiaxion situation, due to axion fields, different from the KR axion, arising from compactification, which depend on the specific (compactified) string model under consideration~\cite{svrcek}.

The existence of periodic modulation potentials for such axions~\cite{sasaki,Mavromatos:2022yql} can also lead to an enhanced production of rotating primordial black holes during inflation, and this can affect the 
duration of the reheating phase~\cite{Mavromatos:2022yql}. 
Under such circumstances and for some regions of the parameter space of the models, there is 
also the possibility of the existence of an intermediate matter-dominated phase, between the RVM inflation and radiation eras~\cite{Carr:2018nkm}. This can affect significantly the populations of primordial black holes, and, as a consequence, the  profiles of the gravitational waves produced from the coalescence of such black holes, leading in turn to observable in principle modifications of the spectrum of the gravitational waves during the early radiation era. In fact, as argued in \cite{Mavromatos:2022yql}, by looking at the details of such gravitational-wave profiles via future interferometers, one can in principle distinguish the effects of the StRVM from generic string-inspired axion-monodromy inflationary models, with linear axion potentials, such as those discussed in \cite{sasaki}. 
A detailed analysis, therefore, of the exit phase from RVM inflation in our StRVM cosmology framework, and a study of the potential  (rotating) primordial-black-hole populations, constitutes an important, potentially very interesting from a phenomenological point of view, avenue for research, that we indent to pursue in the future. 

\acknowledgments

The work of P.D. is supported by a graduate scholarship from the
National Technical University of Athens. The work of N.E.M. is supported in part by the UK Science and Technology
Facilities research Council (STFC) and UK Engineering and Physical Sciences Research Council (EPSRC) under
the research grants  ST/X000753/1 and EP/V002821/1, respectively.
NEM also acknowledges participation in the COST Association Action CA21136 “Addressing observational tensions in cosmology with systematics and fundamental physics (CosmoVerse)”.\\

\section*{APPENDICES} 
\appendix
\section{Derivation of the Equations of the Expansion Normalized (EN) variables}\label{appendixA}

In this Appendix we review the basics of the dynamical system approach to single-scalar-field cosmology. 
In the main text  we have introduced the EN variables  
in Eq.~\eqref{ENvariables},
which we repeat here for the reader's convenience: 
\begin{equation}
    x=\frac{\kappa \dot{b}}{\sqrt{6}H} \ \ \text{and} \ \  y=\frac{\kappa\sqrt{\abs{V}}}{\sqrt{3}H}\,.
    \label{ENVariables}
\end{equation}
Squaring eq.~\eqref{ENVariables}, we obtain: 
\begin{equation}
   \dot{b}^2= \frac{6x^2 H^2}{\kappa^2} \ \ \text{and} \ \  \abs{V}=\frac{3y^2 H^2}{\kappa^2}\,,
\end{equation}
which, by a simple substitution to the Friedmann equation \eqref{freidmann1}, yields:
\begin{equation}
    x^2 + y^2 = 1\,.
\end{equation}
Now, in order to derive the equations for $x^{\prime}$ and $y^{\prime}$,
where the prime denotes differentiation with respect to the quantity $N=\log\alpha$, where $\alpha$ is the scale factor of the Universe, 
we use eq. \eqref{friedmann2}, and obtain:
\begin{align}
      2\dot{H} + 3H^2 = -\kappa^2\left(\frac{\dot{b}^2}{2}-V(b)\right) \quad \Rightarrow \quad
      \frac{\dot{H}}{H^2}=-\frac{3}{2}\left(x^2 -y^2 +1\right)\,.
      \label{friedman_}
\end{align}
We now consider the derivative of $x$ and $y$ with respect to $N=\log\alpha$: 
\begin{equation}
    x^{\prime}\equiv\frac{dx}{dN}=\frac{1}{H}\frac{dx}{dt}=\frac{1}{H}\left(\frac{\kappa \ddot{b}}{\sqrt{6}H} - \frac{\kappa \dot{b}\dot{H}}{\sqrt{6}H^2}\right)=\frac{\kappa \dot{b}}{\sqrt{6}H}\left(\frac{\ddot{b}}{\dot{b}H} - \frac{\dot{H}}{H^2}\right)\,.
    \label{x_prime}
\end{equation}
The two terms inside the brackets on the right-hand side of the last equality in \eqref{x_prime}
can be evaluated by  using the Klein-Gordon equation for the axion \eqref{KG_frw} and also using \eqref{friedman_}. From the equation of motion for the axion, we have:
\begin{align*}
    \ddot{b} + 3H \dot{b}=-V_{,b} \Rightarrow \frac{\ddot{b}}{\dot{b}H}=-3 -\frac{V_{,b}} {\dot{b}H}\,.
\end{align*}
Using the EN variables definitions \eqref{ENVariables} we get:
\begin{equation}
    \frac{\ddot{b}}{\dot{b}H} = -3 - \frac{V_{,b}\sqrt{6}y^2}{2x\kappa V}\,,
    \label{eom_EN}
\end{equation}
and substituting eqs.\eqref{friedman_} and \eqref{eom_EN} into \eqref{x_prime}, we obtain the equation for $x^\prime$:
\begin{equation}
    x^{\prime}=-\frac{3}{2}\left[2x - x^3 +x\left(y^2-1\right)-\frac{\sqrt{2}}{\sqrt{3}}\lambda y^2\right] \,,
\end{equation}
where $ \lambda=-\frac{V_{,b}}{\kappa V}$. This is eq. \eqref{dyn1} for the first dynamical variable that was mentioned in section 
\ref{sec:EN} of 
the main text. Similar steps are followed for the derivation of the  $y^{\prime}$ equation:
\begin{equation}
    y^{\prime}\equiv\frac{dy}{dN}=\frac{1}{H}\frac{dy}{dt}=\frac{\kappa}{\sqrt{3}H}\left(\frac{\dot{V}\abs{V}^{-\frac{1}{2}}}{H} - \frac{\dot{H}\abs{V}^{\frac{1}{2}}}{H}\right)
    \label{y_prime}\,.
\end{equation}
On using:
\begin{align*}
    \dot{V}=\frac{dV}{dt}=\dot{b} V_{,b}=V_{,b}\frac{x \sqrt{6}H}{\kappa}\,,
\end{align*}
eq.~\eqref{y_prime} becomes:
\begin{equation*}
    y^{\prime}=\frac{\sqrt{6}}{2}\frac{V_{,b}}{\kappa V}xy - y\frac{\dot{H}}{H^2}\,,
\end{equation*}
from which, upon substituting  the definition of  $ \lambda=-\frac{V_{,b}}{\kappa V}$ in the first term on the right-hand side, 
and making use of eq~\eqref{friedman_} in the second, we obtain:
\begin{equation}
     y^{\prime}=-\frac{3}{2}y\left[-x^2 + y^2-1 + \frac{\sqrt{2}}{\sqrt{3}}\lambda x\right] \,.
\end{equation}
This is eq.~\eqref{dyn2} in the main text.
The derivation of eq.~\eqref{dyn3} for the evolution of $\lambda$ is straightforward, and is omitted here for brevity.

\section{Stability Analysis of the dynamical system \eqref{phi_prime},\eqref{z_prime}}\label{appendixB}

In this Appendix we will examine the stability of the critical (fixed) points of our dynamical system. In order to study the stability of the fixed points, we first calculate the Jacobian (stability) matrix, which allows us to determine the eigenvalues at each fixed point separately. For the hyperbolic fixed points, the stability can be determined immediately from the eigenvalues, which in such a case should have non-zero real part. Otherwise, if one eigenvalue is zero, corresponding to a center space, more advanced techniques such as center manifold theory or Lyapunov function analysis should be applied in order to draw safe conclusions on the stability of the fixed points. 

For the reader's 
 convenience, 
we rewrite the dynamical system as follows: 
\begin{align*}
        \varphi^\prime & \equiv \mu(\phi,z)=\left( 3\cos\varphi -\frac{\sqrt{6}}{2}\frac{z}{1-z}   \right) \sin\varphi\,,
      \\
      z^\prime & \equiv \nu(\phi,z)=\sqrt{6}z^2 \cos\varphi\,.
    \end{align*}
The linearised system yields the Jacobian (stability) matrix:
\begin{equation}
    J= \begin{pmatrix}
        \frac{\partial \mu}{\partial \phi} &  \frac{\partial \mu}{\partial z} \\
        \frac{\partial \nu}{\partial \phi} &  \frac{\partial \nu}{\partial z}     
        \end{pmatrix}=\begin{pmatrix}
        3 \cos (2 \varphi )+\frac{\sqrt{\frac{3}{2}} z \cos (\varphi )}{z-1} & -\frac{\sqrt{\frac{3}{2}} \sin (\varphi )}{(z-1)^2}\\
       -\sqrt{6} z^2 \sin (\varphi ) & 2 \sqrt{6} z \cos (\varphi )
        \end{pmatrix}\,.
    \label{Jacobian}
\end{equation}
Now, we want to find the eigenvalues of the Jacobian matrix at the fixed points. The points and the eigenvalues are given in Table \ref{full_table_dyn_system} in the main text.

The linear approximation fails in the case when a center manifold exists (that is, at least one eigenvalue has zero real part). Below, we shall  first mention some basic facts about center manifold theory, and then proceed to the analysis for the fixed points.

\begin{center}
    \textit{Center Manifold Theory}
\end{center}
As we have already stated, using linear stability theory, we cannot conclude about the stability of fixed points with eigenvalues that have zero real parts. With the methods of center manifold theory, however, we are able to reduce the dimensions of the dynamical system and then, the stability properties of this reduced system can be investigated analytically. We mention below some of the basics of this method \cite{Bahamonde_2018} and proceed with the analysis for the specific critical points of our dynamical system \eqref{phi_prime},\eqref{z_prime}. 

Let one consider a dynamical system of the following form:
\begin{equation}
    \mathbf{\dot{z}=F(z)}\,,
\end{equation}
where $\mathbf{F}$ is a regular function of $z \in  \mathbb{R}^n$. Assuming a fixed point at $\mathbf{z=z_0}$, we can linearise the system around this point using the Jacobian matrix. Defining $\mathbf{z_*=z-z_0}$, we have that:
\begin{equation}
    \mathbf{\dot{z_*}}=J \mathbf{z_*}
    \label{center_manifold_system}\,.
\end{equation}
We know that the Jacobian $J$ is an $n \times n$ matrix (where $n$ denotes the number of dimensions of the dynamical system), having $n$ eigenvalues. These eigenvalues correspond to three subspaces spanned by the respective eigenvectors. So, $\mathbb{R}^n$ can be expressed as a direct sum of these tree subspaces, denoted by $\mathbb{E}^s,\mathbb{E}^u,\mathbb{E}^c$, with $(s)$ for "stable", $(u)$ for "unstable" and $(c)$ for "center". From the linear stability theory, it is well known that, if all the eigenvalues have positive real parts, the fixed point is \textit{unstable}, while if all the eigenvalues have negative real parts, then the point is \textit{stable}. Mixed signs of the eigenvalues (non-zero) correspond to \textit{saddle} points. So, the eigenvectors of $J$ corresponding to negative real part eigenvalues span the stable subspace $\mathbb{E}^s$ and eigenvectors corresponding to positive real part eigenvalues span the unstable subspace $\mathbb{E}^u$. $\mathbb{E}^c$ is spanned by the Jacobian eigenvectors that are associated to zero real part eigenvalues. The dynamics and the stability properties of the phase space trajectories in both $\mathbb{E}^s$ and $\mathbb{E}^u$ can be analysed  using linear stability theory, while the dynamics in $\mathbb{E}^c$ require the use of center manifold theory to be fully understood.

If the Jacobian has (at least) one eigenvalue with positive real part, then the corresponding fixed point cannot be stable, regardless of it being hyperbolic or non-hyperbolic. Otherwise, there always exists an appropriate coordinate transformation allowing us to rewrite the system \eqref{center_manifold_system} in the form:
\begin{align}
    \mathbf{\dot{x}}&=A \mathbf{x} + f(\mathbf{x,y})
    \label{diagonal_system1}\,,
         \\ 
    \mathbf{\dot{y}}&=B \mathbf{y} + g(\mathbf{x,y}) \,,
    \label{diagonal_system2}
\end{align}
where $(x, y) \in   \mathbb{R}^c \times \mathbb{R}^s$, with $c$ the dimension of the center manifold $\mathbb{E}^c$ and $s$ the dimension of the stable manifold $\mathbb{E}^s$. The functions $f$ and $g$ satisfy:
\begin{align}
    f(0,0)=0 , & \ \ \nabla f(0,0)=0\,,
    \label{f_function}
     \\ 
    g(0,0)=0 , & \ \ \nabla g(0,0)=0  \,.
    \label{g_function}
\end{align}
In the above system \eqref{diagonal_system1},\eqref{diagonal_system2}, $A$ is a $c \times c$ matrix having zero real parts eigenvalues (center manifold), while $B$ is an $s\times s$ matrix having negative real part eigenvalues (stable manifold).
To present the method of center manifold, we have to state some theorems without proof and some definitions, which will then use in order to determine the stability properties of our system. For details
of the analysis, including  proofs of the pertinent theorems, 
the interested reader is referred to Jack Carr's work in \cite{carr1981}.

\textbf{\textit{Definition}:}  A geometrical space is a center manifold for \eqref{diagonal_system1},\eqref{diagonal_system2} if it can be locally represented as
\begin{equation}
    W^{c}(0)=\left[(\mathbf{x,y}) \in \mathbb{R}^{c} \times \mathbb{R}^{s} \  | \ \mathbf{y}=h(\mathbf{x}),\abs{x}<\delta, h(0)=0, \nabla h(0)=0 \right]\,,
    \label{def_center_manifold}
\end{equation}
for $\delta$ being sufficiently small and $h(x)$ a regular function on space $\mathbb{R}^s$.

Next, we present three basic theorems~\cite{carr1981}, which constitute the basis of center manifold theory, allowing us to determine the stability of the fixed points with zero eigenvalues for our cosmological system \eqref{phi_prime}, \eqref{z_prime}.  

\textbf{\textit{Theorem 1}:}  There exists a center manifold for \eqref{diagonal_system1},\eqref{diagonal_system2}, whose dynamics, restricted to the center manifold, are given by
\begin{equation}
    \mathbf{\dot{u}}=A \mathbf{u}+f(\mathbf{u},h(\mathbf{u}))\,,
    \label{difeq_u}
\end{equation}
with $u \in \mathbb{R}^c$ being sufficiently small.

\textbf{\textit{Theorem 2}:} Suppose that the zero solution of \eqref{difeq_u} is stable (or unstable). Then, the zero solution of \eqref{diagonal_system1},\eqref{diagonal_system2} is also stable (or unstable). What is more, if $(\mathbf{x}(t),\mathbf{y}(t))$ is also a solution of \eqref{diagonal_system1},\eqref{diagonal_system2} with $(\mathbf{x}(0),\mathbf{y}(0))$ being small enough, there exists a solution $\mathbf{u}(t)$ of \eqref{difeq_u} such that 
\begin{align}
    \mathbf{x}(t)&=\mathbf{u}(t)+\mathcal{O}(e^{-\gamma t})\,,
    \\
    \mathbf{y}(t)&=h(\mathbf{u}(t))+\mathcal{O}(e^{-\gamma t})\,,
\end{align}
as $t \to \infty $, with the positive constant $\gamma>0$.

These two theorems assume the knowledge of the function $h(\mathbf{x})$ which needs to be calculated. We can derive a differential equation for $h(\mathbf{x})$, using Definition \eqref{def_center_manifold}. We then have that $\mathbf{y} = h(\mathbf{x})$. Differentiating with respect to our time variable, and applying the chain rule, we obtain:
\begin{equation}
    \mathbf{\dot{y}}=\nabla h(\mathbf{x}) \cdot \mathbf{\dot{x}} \,,
\end{equation}
Hence, from \eqref{diagonal_system1},\eqref{diagonal_system2}, we can substitute for $\mathbf{\dot{x}}$ and $\mathbf{\dot{y}}$, and using the fact that $\mathbf{y}=h(\mathbf{x})$, we can derive the following differential equation: 
\begin{equation}
   \mathcal{N}(h(\mathbf{x})):= \nabla h(\mathbf{x}) \left[A \mathbf{x} + f(\mathbf{x},h(\mathbf{x})) \right] - B h(\mathbf{x}) - g(\mathbf{x},h(\mathbf{x}))=0 \,.
    \label{difeq_h}
\end{equation}
The differential equation \eqref{difeq_h} must be satisfied by $h(\mathbf{x})$ in order for it to be the center manifold. In order to approximate the solution of the above equation, which in general is not solvable even in the simplest cases, we present the third theorem:

\textbf{\textit{Theorem 3}:} Let $\phi : \mathbb{R}^c \to \mathbb{R}^s$ be a map with $\phi(0)=\nabla \phi (0)=0$ such that $ \mathcal{N}(\phi(\mathbf{x}))=\mathcal{O}(\abs{\mathbf{x}}^q)$ as $\mathbf{x}\to 0$ for $q>1$. Then, we have that:
\begin{equation}
    \abs{h(\mathbf{x})-\phi(\mathbf{x})}=\mathcal{O}(\abs{\mathbf{x}}^q) \,\, \text{as} \,\, \mathbf{x}\to 0 \,.
\end{equation}
The important thing here is that an approximate description of the center manifold can return the same information about the stability of our fixed point as an exact solution of \eqref{difeq_h}. This can usually be done by a series expansion of $h(\mathbf{x})$, where the coefficients of the series can be determined by satisfying \eqref{difeq_h} for each order. 

Now, we have everything we need to apply the center manifold theory to our system \eqref{phi_prime},\eqref{z_prime} for the fixed points that contain a zero eigenvalue, which are $I(\frac{\pi}{2},0),O(0,0),C(\pi,0)$. 

Let us start with the fixed point of interest, $I(\frac{\pi}{2},0)$. The first step is to "move" the point, via a coordinate transformation, to the origin. So, we send $\phi \to \theta + \frac{\pi}{2}$, and in the new coordinates $\theta,\zeta$, the system \eqref{phi_prime},\eqref{z_prime} becomes: 
\begin{align}
    \theta^{\prime}& =\left( 3\cos\left(\theta+\frac{\pi}{2}\right) -\frac{\sqrt{6}}{2}\frac{\zeta}{1-\zeta}   \right) \sin\left(\theta+\frac{\pi}{2}\right)\,,
    \label{theta_prime}
    \\
    \zeta^{\prime}& = \sqrt{6}\,\zeta^2 \cos\left(\theta+\frac{\pi}{2}\right)\,,
    \label{z_prime2}
\end{align}
where now the fixed point $I(\phi=\frac{\pi}{2},\zeta=0)$ corresponds to $I(\theta=0,\zeta=0)$. The Jacobian matrix of the system \eqref{theta_prime}, \eqref{z_prime2} reads: 
\begin{equation}
    J =\begin{pmatrix}
        -3 \cos (2 \theta )-\frac{\sqrt{\frac{3}{2}} \zeta \sin (\theta )}{\zeta-1} & -\frac{\sqrt{\frac{3}{2}} \cos (\theta )}{(\zeta-1)^2} \\
        -\sqrt{6} \zeta^2 \cos (\theta ) &  -2 \sqrt{6} \zeta \sin (\theta )
        \end{pmatrix}\,,
    \label{Jacobian2}
\end{equation}
and, at $I(\theta=0,\zeta=0)$, we have:
\begin{equation}
    J |^{\theta=0}_{\zeta=0} =\begin{pmatrix}
        -3 &  -\sqrt{\frac{3}{2}} \\
        0 & 0
        \end{pmatrix}\,,
    \label{Jacobian_I}
\end{equation}
with the eigenvalues and the coressponding eigenvectors given by:
\begin{align}
 &\lambda_1=0 \ \to \ \mathbf{u}=\left(\begin{array}{c}
 -\frac{1}{\sqrt{6}} \\
 1 \\
\end{array}\right)\,,
\\
 &\lambda_2=-3 \ \to \ \mathbf{v}=\left(\begin{array}{c}
 1 \\
 0 \\
\end{array}\right)\,.
\end{align}
The eigenvalue $\lambda_2 = -3$ and the corresponding eigenvector $\mathbf{v}$ span the stable subspace for the fixed point, while the eigenvalue $\lambda_1 = 0$ and the corresponding eigenvector $\mathbf{u}$ span the center subspace. The next step is to find new variables, which diagonalize the Jacobian matrix \eqref{Jacobian_I}, in order to obtain the necessary form of eq. \eqref{diagonal_system1},\eqref{diagonal_system2} so as to be able to apply the arguments of the center manifold theory. 

To this end, we start with the linearized system 
\begin{equation}
    \left(\begin{array}{c}
 \theta^{\prime} \\
 \zeta^\prime \\
\end{array}\right)=  J |^{\theta=0}_{\zeta=0} \left(\begin{array}{c}
 \theta \\
 \zeta \\
\end{array}\right)\,,
\label{linear_theta_z}
\end{equation}
and the Jacobian 
\begin{align*}
    J=P D P^{-1} \,,
\end{align*}
where
\begin{equation}  
    P =\begin{pmatrix}
        -\frac{1}{\sqrt{6}}  & 1\\
        1 & 0
        \end{pmatrix} , \ P^{-1} =\begin{pmatrix}
        0 &  1\\
        1 & \frac{1}{\sqrt{6}}
        \end{pmatrix}  , \  D =\begin{pmatrix}
        0 &  0\\
        0 & -3
        \end{pmatrix} 
\end{equation}
are the matrices from the diagonilisation of the Jacobian matrix at 
the fixed point $I(\theta=0,z=0)$. Now, multiplying \eqref{linear_theta_z} with $P^{-1}$ from the left on both sides, and keeping in mind that $D=P^{-1} J P$, we get:
\begin{equation}
    P^{-1} \left(\begin{array}{c}
 \theta^{\prime} \\
 \zeta^\prime \\
\end{array}\right)= D P^{-1} \left(\begin{array}{c}
 \theta \\
 \zeta \\
\end{array}\right)\,.
\label{diagonilise_coords}
\end{equation}
So, an appropriate (diagonal) coordinate transformation is
\begin{equation}
     \left(\begin{array}{c}
 U \\
 V \\
\end{array}\right)=P^{-1} \left(\begin{array}{c}
 \theta \\
 \zeta \\
\end{array}\right)=\begin{pmatrix}
        0 &  1\\
        1 & \frac{1}{\sqrt{6}}
        \end{pmatrix} \left(\begin{array}{c}
 \theta \\
 \zeta \\
\end{array}\right)\,,
\label{coord_U_V}
\end{equation}
which yields
\begin{equation}
    U=\zeta \ \  \text{and} \ \  V=\theta+\frac{\zeta}{\sqrt{6}}\,.
\end{equation}
Eq. \eqref{diagonilise_coords} can be written as 
\begin{equation}
   \left(\begin{array}{c}
 U^\prime \\
 V^\prime \\
\end{array}\right) = D \left(\begin{array}{c}
 U \\
 V \\
\end{array}\right)=\begin{pmatrix}
        0 &  0\\
        0 & -3
        \end{pmatrix} \left(\begin{array}{c}
 U \\
 V \\
\end{array}\right)\,,
\end{equation}
which can be cast in the form of eqs.~\eqref{diagonal_system1}, \eqref{diagonal_system2}
\begin{align}
    U^\prime &= 0 \,,\\ 
    V^\prime &= -3 V\,,
\end{align}
giving the values of $A=0$ and $B=-3$ (see \eqref{diagonal_system1}, \eqref{diagonal_system2}). Now, we can express the system of \eqref{theta_prime}, \eqref{z_prime2} in terms of the the new variables $U$ and $V$, which reads:
\begin{align}
   U^\prime &= A U + f(U,V) \,,\\  
   V^\prime &= B V + g(U,V)\,,
\end{align}
where $A=0,B=-3$ and the functions $f,g$ are given by:
\begin{align}
    f(U,V) &= \sqrt{6}\sin\left(\frac{U}{\sqrt{6}}-V\right)U^2 \,,
    \label{f(u_v)}
    \\ 
    g(U,V) &=  U^2 \sin \left(\frac{U}{\sqrt{6}}-V\right)-\frac{6 (U-1) \sin \left(\frac{U}{\sqrt{6}}-V\right)+\sqrt{6} U }{2 (1-U)}\cos \left(\frac{U}{\sqrt{6}}-V\right)+3 V\,.
    \label{g(u_v)}
\end{align}
These functions indeed satisfy eqs~\eqref{f_function}, \eqref{g_function}:
    \begin{align}
    f(U=0,V=0)=0 , & \ \ \nabla f(U=0,V=0)=0\,,
     \\ 
    g(U=0,V=0)=0 , & \ \ \nabla g(U=0,V=0)=0  \,,
\end{align}
which is an important feature, allowing us to proceed to the description of the center manifold stability properties. 

To this end, we have to find an approximate solution for the function $h(U)$ for our center manifold. We can form the differential equation \eqref{difeq_h}, with $A=0,B=-3$ and $g(U,V)$ given from \eqref{g(u_v)}, where $V$ is a function of U, $V=h(U)$ locally. Then, we have the differential equation: 
\begin{align}
    &h^{\prime}(U) \left[A U + f(U,h(U)) \right] - B h(U) - g(U,h(U))=0 \quad \Rightarrow 
    \quad  
     &h^{\prime}(U)f(U,h(U))+3h(U)-g(U,h(U))=0\,.
\end{align}
We can solve the above equation for the first terms of a series expansion in $h(U)$. Keeping the leading orders, with the boundary condition $h(0)=0$, gives: 
\begin{equation}
    h(U)=-\frac{U^2}{\sqrt{6}}-\frac{25 U^3}{36 \sqrt{6}}-\frac{U^4}{12 \sqrt{6}} +\mathcal{O}(U^5)\,.
\end{equation}
Now that we have found an approximate solution for $h(U)$, we can reduce the dimensionality of the problem, and define the stability of our dynamical system \eqref{phi_prime},\eqref{z_prime} only by studying the stability properties of eq. \eqref{difeq_u}. Thus, we have:
\begin{align}
   u^{\prime}(N)&=A u(N) + f(u(N),h(u(N)))\,,
\end{align}
which gives 
\begin{equation}
     u^{\prime}(N)=\sqrt{6} u(N)^2 \sin \left(\frac{u(N) \left(-677 u(N)^4+120 u(N)^3+1000 u(N)^2+1440 u(N)+1440\right)}{1440 \sqrt{6}}\right)\,.
\end{equation}
Upon a series expansion of the right hand side of the above equation, we obtain:
\begin{equation}
     u^{\prime}(N) = u(N)^3-\frac{u(N)^5}{3}+\mathcal{O}\left(u(N)^6\right)\,.
\end{equation}
The sign of the constant in front of the term $u(N)^3$ gives us all the information we need about the stability of the direction related to the zero eigenvalue of our fixed point $I(\phi=\frac{\pi}{2},\zeta=0)$. Positive coefficient means instability along the eigenvector of the zero eigenvalue, while a negative one implies stability. In our case, the coefficient is $+1>0$, so we have an unstable direction concerning the center manifold. This means that $I(\phi=\frac{\pi}{2},\zeta=0)$ is a saddle point, with one stable $(\lambda=-3)$ and one unstable $(\lambda=0)$ direction. 

Similar analysis can be done for every fixed point with a zero eigenvalue. The complete results are summarized in Table \ref{full_table_dyn_system} in section \ref{sec:EN} of the main text.

\newpage

\bibliography{bibliographyDMV}

\end{document}